\documentclass[usenatbib]{mnras}

\usepackage{newtxtext,newtxmath}
\usepackage{placeins}

\usepackage[T1]{fontenc}
\usepackage{euscript}
\usepackage{booktabs}
\DeclareRobustCommand{\VAN}[3]{#2}
\let\VANthebibliography\thebibliography
\def\thebibliography{\DeclareRobustCommand{\VAN}[3]{##3}\VANthebibliography}

\usepackage{ulem}
\usepackage{subcaption}
\usepackage{tablefootnote}

\usepackage[dvipsnames]{xcolor}
\usepackage{comment}
\usepackage{graphicx}	
\usepackage{amsmath}	
\usepackage{hyperref}
\usepackage{ulem,xcolor}

\usepackage{multirow}

\usepackage{enumitem}

\usepackage{soul,color}

\defcitealias{Varadaraj_2023}{V23}

\defcitealias{2000ApJ...539..718C}{CF00}

\defcitealias{2020MNRAS.495.4040W}{W20}

\title[The host galaxies of SNe Ia]{The dependence of the Type Ia Supernova colour--luminosity relation on their host galaxy properties}

\author[S. Ramaiya et al.]{
S. Ramaiya,$^{1}$\thanks{E-mail: shruti.ramaiya@physics.ox.ac.uk}
M. Vincenzi,$^{1}$
M. J. Jarvis,$^{1,2}$
P. Wiseman,$^{3}$
M. Sullivan$^{3}$
\\
$^{1}$Astrophysics, Department of Physics, University of Oxford, Keble Road, Oxford, OX1 3RH, UK\\
$^{2}$Department of Physics and Astronomy, University of the Western Cape, Robert Sobukwe Road, 7535 Bellville, Cape Town, South Africa\\
$^{3}$School of Physics and Astronomy, University of Southampton, Southampton, SO17 1BJ, UK
}

\date{Accepted XXX. Received YYY; in original form ZZZ}

\pubyear{\the\year{}}

\begin{document}
\label{firstpage}
\pagerange{\pageref{firstpage}--\pageref{lastpage}}
\maketitle

\begin{abstract}

Using the Dark Energy Survey 5-year sample, we determine the properties of type Ia supernova (SN Ia) host galaxies across a wide multi-wavelength range -- from the optical to far-infrared -- including data from the \textit{Herschel} and \textit{Spitzer} space telescopes. We categorise the SNe Ia into three distinct groups according to the distribution of their host galaxies on the star-formation rate (SFR) -- stellar mass ($M_\star$) plane. Each region comprises host galaxies at distinct stages in their evolutionary pathways: Region 1 -- low-mass hosts; Region 2 -- high-mass, star-forming hosts and Region 3 -- high-mass, passive hosts. We find SNe Ia in host galaxies located in Region 1 have the steepest slope (quantified by $\beta$) between their colours and luminosities, with $\beta_{\mathrm{R_1}} = 3.51 \pm 0.16$. This differs at the ${\sim}6\sigma$ significance level to SNe Ia in Region 3, which have the shallowest colour--luminosity slope with $\beta_{\mathrm{R_3}} = 2.12 \pm 0.16$. After correcting SNe Ia in each subsample by their respective $\beta$, events in Region 3 (high-mass, passive hosts) are $0.07 - 0.12$ mag ($>3\sigma$) brighter, post-standardisation. 
We conclude that future cosmological analyses  should apply standardisation relations to SNe Ia based upon the region in which the SN host galaxy lies in the SFR--$M_\star$ plane. Alternatively, cosmological analyses should restrict the SN Ia sample to events whose host galaxies occupy a single region of this plane.

\end{abstract}




\section{Introduction}

Owing to the discovery of the standardizable nature of type Ia supernovae (SNe Ia), today, they stand among the best distance indicators, with estimates to within a ${\sim}7\%$ accuracy ($\simeq0.14$ mag; \citealp{2018ApJ...859..101S}), out to redshifts, $z \sim 1$. In the decades following the first detection of the accelerating expansion of the Universe (\citealp{1998AJ....116.1009R, Perlmutter_1999}), the use of SNe Ia as cosmological tracers has motivated the launch of new and more powerful time-domain surveys, to observe and follow up increasing numbers of transients. On account of the growing sample size of well-observed SNe Ia, and the increasingly precise statistical constraints they provide on dark energy, we now enter an era where the accuracy of any future cosmological analyses is governed by systematic sources of error \citep[e.g.,][]{2006A&A...447...31A, 2009ApJS..185...32K, 2009ApJ...700.1097H, 2011ApJS..192....1C, 2011ApJ...737..102S, 2014A&A...568A..22B}. Historically, the leading systematic in analyses of SNe Ia has been photometric calibration. More recently, studies \citep[e.g.,][]{ 2019ApJ...874..150B,2022ApJ...938..110B, 2024ApJ...975...86V} have begun to indicate that understanding the intrinsic scatter that remains after the standardisation of SN Ia luminosities  constitutes a source of systematic uncertainty of comparable scale.

Current efforts to standardize SN Ia luminosities are conducted on a purely empirical basis and are built upon  a two-parameter correction. The first is to the light curve decline rate (hereafter `stretch'): where SNe with slower declining, broader light curves are brighter and those with more rapidly evolving, narrower light-curves are fainter \citep{1974PhDT.........7R,1977SvA....21..675P,1993ApJ...413L.105P}. This relation is believed to be the result of the mass of$\ ^{56}$Ni synthesised in the explosion \citep{1982ApJ...253..785A}. The second parameter is colour \citep[e.g.,][]{1996ApJ...473...88R, 1998A&A...331..815T, Wang_2003} with brighter SNe being bluer and their fainter counterparts, redder. The origin of this variation in colour is poorly understood but is believed to result from a combination of the intrinsic SN colour and intervening dust along the line-of-sight. When combined, accounting for these two relations reduces the observed dispersion in SN peak magnitudes from ${\sim}0.35$ mag  to ${\sim}0.14$ mag \citep{2018ApJ...859..101S}. As for the `intrinsic scatter' that remains (after accounting for observational errors), its origin, while yet unknown, may arise from uncertainties attributed to a number of sources e.g., explosion physics, progenitor configuration -- both of which link to the focus of this work, the host galaxies of SNe Ia.

It is well established that the observed properties of SNe Ia are influenced by the physical characteristics of their host galaxy environments. Much of the earlier work, driven by the observation that SNe Ia are discovered in all types of galaxies, has focused on relations between SN properties and their host morphologies. For instance, the optical light-curves of SNe Ia in elliptical galaxies exhibit a narrower dispersion in their decline rate \citep{1989PASP..101..588F, 1996AJ....112.2391H}. SNe Ia in morphologically early-type ellipticals (and/or passive environments) are also fainter and possess more rapidly-declining light curves than systems with active, ongoing star-formation, such as spiral galaxies; which preferentially host the longest lived, brightest SNe Ia \citep{1999AJ....117..707R,2000AJ....120.1479H,2001ApJ...554L.193H, 2006ApJ...648..868S}.

Studies concerning the link between SN Ia properties and host galaxy types remain an ongoing area of research. More recent efforts in the field, however, have begun shifting their focus to attempting to understand the relation between SN luminosities post-standardisation and their host properties. Of the most notable (or at least, most widely discussed) is the correlation with galaxy stellar mass \citep{2010ApJ...715..743K,2010MNRAS.406..782S,2010ApJ...722..566L}, where colour- and stretch-corrected SNe Ia in high mass ($>$10$^{10}M_\odot$) galaxies are observed to be ${\sim}0.06 - 0.15$ mag brighter, on average, than those in low mass galaxies. It has therefore become commonplace in cosmological analyses to apply a third ad-hoc correction during the standardisation of SN Ia luminosities. This is usually modelled via a step function, where SNe in galaxies on either side of this 10$^{10}$M$_\odot$ threshold are corrected by different values (aptly named the \lq mass step\rq). Note, that whilst stellar mass is the preferred host galaxy parameter used when standardising SN Ia luminosities, trends have also been identified with global and local\footnote{`Global' refers to entire galaxy measurements and `local' refers to measurements in $\sim 1-4$kpc radius regions centered on the SN location.} specific star formation rate \citep[sSFR;][]{2010MNRAS.406..782S,2020A&A...644A.176R}, galaxy colour \citep{2018A&A...615A..68R, 2021MNRAS.501.4861K, 2023MNRAS.519.3046K}, average stellar population age \citep{2011ApJ...740...92G,2019ApJ...874...32R} and gas-phase metallicity \citep{2011ApJ...743..172D,2013ApJ...770..108C}, to name only a few. 

The physical underpinnings for the aforementioned trends remain, as yet, largely unknown. For the \lq mass step\rq\ in particular, various hypotheses have been considered in the literature. Some studies suggest it may arise due to the metallicity of the underlying stellar populations \citep{2010MNRAS.406..782S}. The metallicity (both gas-phase and stellar) of a galaxy is contingent upon the depth of its gravitational potential well: higher mass galaxies are able to hold onto a larger proportion of the metals they synthesise, due to having deeper potential wells, whereas the shallower potential wells of low mass galaxies means they are more susceptible to losing metals via mechanisms such as galactic winds and outflows \citep{2004ApJ...613..898T, 2005MNRAS.362...41G,2010MNRAS.409..421G}. Others explore the origin of the step in relation to the effects of progenitor age; which may correlate with the cumulative age of the stellar population and therefore, galaxy mass \citep{2013ApJ...770..108C,2013A&A...560A..66R}. A third possibility that has inspired more recent works (e.g, \citealp{2021ApJ...909...26B}) is that the mass step is a consequence of variations in the average dust properties between low and high mass galaxies. This is partly motivated from the observation that galaxies exhibit a large range of dust extinction laws, which are usually parametrised by the total-to-selective extinction parameter, $R_V = A_V/(A_B - A_V)$, where $A_V$ and $A_B$ are the attenuation\footnote{"Extinction" and "attenuation" are often used interchangeably in the literature and the same applies in this paper. In sections where the conceptual differences between the two terms become important, we note this explicitly.} in the $V$ and $B$ band, respectively ($A_B - A_V$ is commonly referred to as `reddening' and represented by $E(B-V)$ in the literature). For example, studies of galaxies in the local group have been measured to have values in the range of $R_V \sim 2 -5$ \citep{2013ApJ...776....7G, 2017ApJ...847..102Y}, implying the presence of dust grains which differ in grain size and composition. It is worth noting that metallicity, age and dust are all correlated and that the mass step may be a consequence of all three effects. Disentangling the contribution from each is a non-trivial task.

Using a sample of SNe Ia from the Dark Energy Survey (DES), \cite{Meldorf_2022} model the spectral energy distributions (SED) of SN Ia host galaxies, using optical and near-infrared (NIR) data, with the aim of constraining parameters that characterise dust ($A_V$ and $R_V$, to be exact). They find a large distribution of $R_V \sim 1 - 6$ across their host sample and that, on average, the $R_V$ values exhibited by high stellar mass hosts are $\sim 0.7$ lower than their low-mass counterparts. The authors continue by exploring correlations between their global dust parameter constraints and SN properties, revisiting the concept of the "mass step". According to their findings, applying host galaxy dust attenuation ($A_V$) corrections to SN Ia luminosities greatly reduces the significance of the step size to ${\sim }1.4\hspace{0.05cm}\sigma$. \cite{Meldorf_2022} interpret these results on the grounds of previous studies (\citealp{2020ARA&A..58..529S}) that investigate correlations between dust and stellar mass. They suggest that not only are these indications that the origin of the mass step is partly due to dust but that accounting for host galaxy dust has the potential to reduce the intrinsic scatter that remains post-standardisation. A key limitation of the analysis by \cite{Meldorf_2022}, however, is the attempt to constrain host galaxy dust parameters using \textit{only} optical and NIR data. 

The SED of a galaxy is the sum total of all its individual components e.g., stars, dust, gas etc. 
At optical and NIR wavelengths ($\lambda \lesssim 5 \hspace{0.1cm} \mu$m), galaxy spectra are characterised by emission coming mainly from stellar photospheres. At longer wavelengths, contributions from nebular emission, as a result of reprocessed starlight, and the dust emission curve begin to dominate (see \citealp{Conroy_2013} for an excellent review). Therefore, studies that aim to understand the properties of dust in galaxies (in general, not limited to SN hosts) but use only optical/NIR data are probing the part of the SED that is dominated by stellar light, as opposed to dust-dominated wavelengths. In this regime (to reiterate, at $\lambda \lesssim 5 \hspace{0.1cm} \mu$m), dust affects observations in two ways: it acts to both dim and redden spectral light. Therefore, much of the existing literature is based on the premise that the amount of reddening observed at these wavelengths provides an accurate measure of the level of dust obscuration affecting a galaxy's light. To a limited degree, this holds true; however, there are additional parameters characterising a galaxy's SED that make it subject to degeneracies in the optical/NIR. Other factors can mimic the role that dust has of reddening a spectrum such as, the age of the stellar population or metallicity effects (the so called, `age-metallicity-dust' degeneracy; \citealp{1994ApJS...95..107W, 2001ApJ...559..620P}) e.g., a relatively low-attenuation galaxy with old (red) stars may be indistinguishable from a dusty galaxy comprising of a younger, bluer stellar population. This then begs the question of the validity of dust parameter constraints obtained from analyses using limited range, short wavelength ($\lambda \lesssim 5 \hspace{0.1cm} \mu$m) data and brings us to the focus of this work.

In this paper, we aim to constrain dust properties of SN Ia host galaxies by incorporating mid- to far-infrared data obtained from the \textit{Spitzer} and \textit{Herschel} space telescope missions (spanning $\lambda \sim 3.6 - 500$ $\mu$m), in addition to optical and NIR observations. This allows us to study the impact of adding longer wavelength photometric data, not only on dust parameters e.g., $A_V$, but also on other galaxy properties i.e. stellar mass and star formation rate. Using $A_V$ estimates derived when considering the full breadth of the galaxy SED, we explore potential correlations between global host dust attenuation and SN Ia properties. Finally, we use our robust host galaxy parameter estimates to investigate sub-samples of SNe Ia, split according to the region their host galaxies occupy on the star-formation rate -- stellar mass plane.

The structure of this paper is as follows. In Section~\ref{DES-SN} we present the subset of DES from which our SNe Ia originate. In Section~\ref{hostcat} we introduce the multi-wavelength host galaxy imaging catalogues used in our analysis. We step through the various methods used to obtain photometric measurements in the specific wavebands in Section~\ref{phot}, along with details pertaining to the compilation of the final catalogues used in this work. We describe our SED fitting framework in Section~\ref{bag}. The results of this analysis are given in Sections~\ref{results} -- \ref{results3}, followed by a discussion in Section \ref{discussion}. We present our conclusions in Section \ref{conclusion}. 

Throughout the paper, where relevant, we assume a flat $\Lambda$CDM cosmological model with $\Omega_\mathrm{M} = 0.315$, $\Omega_\Lambda = 0.685$ and $\mathrm{H_0} = 70$ km~s$^{-1}$ Mpc$^{-1}$ (following Planck Collaboration VI \citeyear{2020}). 

\section{Dark Energy Survey Supernova sample} \label{DES-SN}

DES is an optical imaging survey that ran for a period of six years and covered {$\sim$}5100 deg$^2$ of the southern hemisphere using the Dark Energy Camera \citep[DECam;][]{2015AJ....150..150F} on the Blanco 4-m telescope at the Cerro Tololo Inter-American Observatory. DES was designed with the purpose of constraining the properties of dark energy using four complementary probes: galaxy clusters, weak gravitational lensing, large scale structure and SNe Ia. The DES Supernova Programme (DES-SN) was the time-domain component of DES designed to detect thousands of cosmologically-useful SN Ia light-curves over a redshift range 0.05 < \textit{z} < 1.2 \citep{2012ApJ...753..152B}. 

DES-SN monitored ten 2.7-deg$^2$ fields in $griz$ in the southern hemisphere over a period of five years. Eight were `shallow' fields with a depth on each epoch of {$\sim$}23.5 mag and two were `deep’ fields, with a depth of {$\sim$}24.5 mag. The locations of the DES-SN fields were chosen to coincide with four well-studied extragalactic legacy fields \citep{2020AJ....160..267S} containing pre-existing (and forthcoming) ancillary multi-wavelength datasets . The regions that the DES-SN footprint overlapped with (and from which the prefix for each DES field name is sourced) were: Chandra Deep Field---South (CDFS; `C1' and `C2' shallow fields, and the `C3' deep field), ELAIS--S1 (ES1; `E1' and `E2' shallow fields), Stripe 82 (`S1' and `S2' shallow fields) and XMM--Large Scale Structure (XMM--LSS; `X1' and `X2' shallow fields, and the `X3' deep field).

DES-SN surveyed each field with a typical $\sim$7-day cadence and discovered over ${\sim}$30,000 SN candidates with a small fraction followed up spectroscopically \citep[e.g.,][]{2020AJ....160..267S}. 
Photometrically classified SNe Ia from the full five years of DES, for which spectroscopic redshifts from the host galaxies are available, form the `DES-SN5YR' sample \citep{2024ApJ...975....5S}. Recent cosmological analyses \citep{2024ApJ...973L..14D} using DES-SN5YR have helped place one of the tightest constraints on the dark energy equation of state parameter, $w$, especially in combination with Baryonic Acoustic Oscillation (BAO) data from the Dark Energy Spectroscopic Instrument (DESI) \citep{Adame_2025}. In this paper we use data from DES-SN5YR. After applying the selection cuts described below, our final sample consists of 1685 SNe Ia. This includes SNe Ia across all ten DES fields. Additional sample cuts are made at later stages in the analysis based on the availability of multi-wavelength host galaxy data (Section \ref{hostcat}).

\subsection{Sample selection}

To select likely SNe Ia from the photometrically-identified DES-SN candidates, we follow the baseline approach adopted in the DES-SN5YR cosmological analysis \citep{2024ApJ...975...86V}. Using probabilities determined with the photometric classifier \texttt{SuperNNova} \citep{2020MNRAS.491.4277M}, we select events with probability of being a SN Ia, $P_{\mathrm{Ia}}$, larger than $0.5$.

The same light-curve quality cuts are applied as in \cite{2024ApJ...975...86V}, with the following exception. We do not apply any cut on the SN colour (typically $-0.3<c<0.3$) and consider the full range of SN colour allowed within the SALT3 framework ($[-0.5, 0.5]$), thus including in our analysis very blue and red SNe.

\subsection{SN Ia distance estimation}

SN Ia distance moduli, $\mu_{\mathrm{obs}}$ are estimated using (e.g., \citealp{1998A&A...331..815T, 2006A&A...447...31A})
\begin{equation}
    \mu_{\mathrm{obs}} = m_x + \alpha x_1 - \beta c - M - \Delta \mu_{\mathrm{bias}} ,
    \label{eq:SNstand}
\end{equation}

where $m_x$, $x_1$ and $c$ are the SN Ia light-curve parameters as defined in the SALT3 model framework \citep{2007A&A...466...11G,2021ApJ...923..265K}, representing the light-curve amplitude, stretch and colour, respectively. The light curve fit parameters used in this analysis come from the DES-SN5YR data release\footnote{\label{desdr}\url{https://github.com/des-science/DES-SN5YR} }\citep{2024ApJ...975....5S}. The nuisance parameters $\alpha$ and $\beta$ parametrise the stretch--luminosity and colour--luminosity relations, respectively, and $M$ is the absolute magnitude of a SN Ia with $x_1=0$ and  $c=0$. Biases arising from various selection effects and choices in analysis are accounted for using the $\Delta \mu_{\mathrm{bias}}$ term. Standard cosmological analyses include an additional term that accounts for any residual dependencies between standardised (i.e., colour- and stretch-corrected) SN Ia luminosities and their host galaxy properties. This is typically denoted by $\gamma G_{\mathrm{host}}$ and defined as a step function of the form,
\begin{equation}
    \gamma G_{\mathrm{host}} = \begin{cases}
        +\gamma/2 & P > P_{\mathrm{step}}, \\
    -\gamma/2 & \mathrm{otherwise}, 
    \end{cases}
\end{equation}
where $\gamma$ is the residual `step' size, $P$ is a chosen property of the SN host galaxy and $P_{\mathrm{step}}$ is the threshold value defining the step. It is usual in cosmological analyses of SNe Ia to take the stellar mass as the host galaxy property, with the step measured on either side of a 10$^{10}M_\odot$ threshold (i.e., the "mass step"). In the interest of studying correlations between SNe Ia and their host galaxies, we do not use this term in our analysis.

\subsection{Sample selection effects and corrections}
\label{BBC}

Like most high-$z$ SN surveys, DES is magnitude-limited. As a result, the DES-SN sample is incomplete at higher redshifts and biased toward the brightest SNe due to the Malmquist bias. Furthermore, the requirement for each DES SN to have a spectroscopic redshift from its host introduces additional selection biases, favouring SNe in brighter host galaxies. DES SN selection biases have been characterized and described in detail \citep{2019MNRAS.485.1171K,2021MNRAS.505.2819V, 2024ApJ...975...86V}. From these analyses, we can estimate that the DES SN sample is approximately complete up to $z \sim 0.4$ in the shallow fields and $z \sim 0.65$ in the deep fields.

We correct for sample selection biases using the "BEAMS with Bias Corrections" \citep[BBC: ][]{KS17} framework. The BBC framework has been used in many recent cosmological analyses, including the DES-SN5YR cosmological analysis, and is implemented to estimate both corrections for selection effects ($\Delta\mu_{\mathrm{bias}}$ in eq.~\ref{eq:SNstand}) and the best-fit SN standardization parameters ($\alpha$, $\beta$ and $\gamma$ in eq.~\ref{eq:SNstand}).

With BBC, model selection biases are modelled using large simulations of the DES-SN sample. We use the DES-SN5YR simulations described in \cite{2021MNRAS.505.2819V}. These simulations incorporate DES observational noise and, most importantly, galaxy-dependent selection effects due to the requirement of a SN host spectroscopic redshift. In this analysis, we model selection bias corrections $\Delta\mu_{\mathrm{bias}}$ as a function of SN redshift only (the so-called BBC "1D" approach). 

SN standardization parameters are usually estimated as the set of parameters that minimize the scatter in the "Hubble residuals", i.e., $\mu_{\mathrm{obs}}-\mu_{\mathrm{cosmo}}$, where $\mu_{\mathrm{cosmo}}$ are the SN Ia distances expected from some reference cosmology. In BBC, we can remove any dependence on the arbitrary chosen reference cosmology by fitting for additional global redshift-dependent offsets \citep[this approach was first introduced by][]{2011ApJ...740...72M}. BBC performs the fitting using the minimization code \texttt{MINUIT}.

In order to test the robustness of our results after selection effects, in our analysis we use both the full DES-SN sample (up to $z \sim$ 1.2) and apply a redshift cut of $z<0.6$ (50\% of the full sample) since, in this redshift range, selection effects are expected to be significantly smaller.

\section{Host Galaxy Datasets} \label{hostcat}

This section provides a summary of the multi-wavelength host galaxy catalogues that overlap with the DES-SN regions, including the surveys and instruments from which they are derived. We note that nominal DES analyses of SN Ia hosts have been performed with deep-stacked DES \textit{griz} photometry using all available SN survey epochs, minus those in a given year to avoid SN light contaminating the hosts (\citealp{2020MNRAS.495.4040W}, hereafter \citetalias{2020MNRAS.495.4040W}). This photometry was supplemented by DES \textit{u} and VISTA \textit{JH$K_s$} (\citealp{2022MNRAS.509.3547H}), where available, and presented for SN hosts by \cite{2023MNRAS.519.3046K}. Here, we aim for the best and largest multi-wavelength spectral coverage and thus only consider SN hosts located across CDFS and XMM because, subject to availability, data across these fields allow us to probe the maximum extent of a galaxy's SED. Therefore, this work only includes four of the eight DES SN shallow fields (`C1', `C2', `X1', `X2') and both deep fields (`C3', `X3'). Future work may extend to include ELAIS--S1, when similar quality optical data has been collated and combined into a consistent matched catalogue. In the following sub-sections, we present the combination $u-$band through to far-infrared data in these fields that we use for our host galaxy SED modelling.

\begin{figure*}
    \centering
    \includegraphics[width = 180mm]{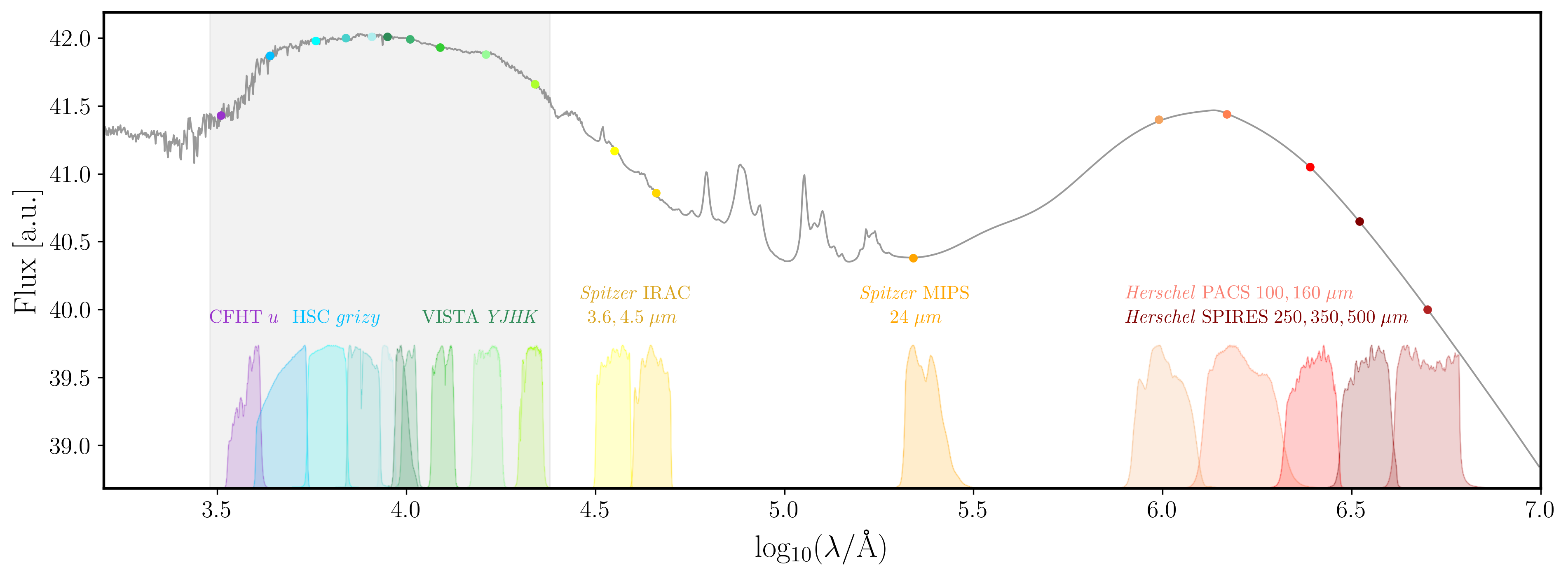}
    \caption{Galaxy SED model superimposed with photometric filters used in this work, with a total of 21 filters. The wavebands span from the \textit{u}$-$band to \textit{Herschel} SPIRE data at 500\,$\mu$m. We exclude the VST \textit{ugri} bands from this Figure as they are complementary to the CFHT \textit{u} and HSC \textit{gri} bands. The grey shaded region indicates the limited range in wavelengths used in many current studies of SN hosts; probing only the part of the SED that is dominated by stellar light ($\lambda \lesssim 5 \mu m$).}
    \label{frank}
\end{figure*}

\subsection{Optical}

\subsubsection{Canada-France-Hawaii Telescope (CFHT) - u$^*$}

The 3.6-m Canada-France-Hawaii Telescope (CFHT) is an optical/infrared telescope located near the summit of Mauna Kea, Hawaii. The CFHT Legacy Survey (CFHT-LS; \citealp{2012SPIE.8448E..0MC}) was a Large Program that ran between 2003 -- 2009. It used the 1-square-degree MegaCam imager, with its 0.18 arcsec per pixel resolution, \citep{2003SPIE.4841...72B} installed at the prime focus MegaPrime environment on the CFHT. More than 450 nights of telescope time were allocated to CFHT-LS, which comprised three survey components: the `Deep', `Wide' and `Very Wide', with observations taken in the $u^*g' r' i' z'$ bands. We take CFHT-$u^*$ band data over XMM-LSS from the $K_{\mathrm{s}}$--band selected version of the catalogues used in \cite{2023MNRAS.523..327A}. For information on field coverage and filter depths, see \cite{2023MNRAS.523..327A}, Fig. 1 and Table 1. CFHT-LS did not cover CDFS. Whilst ancillary data from the CFHT instrument exists over this field, we instead opt for $u$--band data from the VST VOICE survey covering CDFS.

\subsubsection{VLT Survey Telescope (VST)-VOICE - u,g,r,i}

The VST Optical Imaging of the CDFS and Elais-S1 (VOICE; \citealp{vaccari2017voicesurveyvst}) survey was carried out using the OmegaCAM wide-field imager, with its 0.216 arcsec per pixel resolution, on ESO Paranal's 2.6-m VLT Survey Telescope. VOICE covers two extra-galactic regions of the southern hemisphere, with 8 deg$^2$ split equally between CDFS and ES1. Of the 4 deg$^2$ VOICE coverage of CDFS, 3.89 deg$^2$ also contains overlapping data from VISTA and HSC. We take VOICE \textit{ugri} band data from catalogues created by \citet[hereafter \citetalias{Varadaraj_2023}]{Varadaraj_2023}, for which the 5$\sigma$ limiting magnitudes are 25.5, 26.0, 26.0 and 24.7 mag in each of the respective bands. The VOICE data is used to complement the HSC imaging across CDFS. The reasons for this are described in \citetalias{Varadaraj_2023}, Section 2.2 but, in short, both HSC-\textit{I} and -\textit{Z} are affected by poor seeing due to a range of observing conditions, whilst HSC-\textit{R} has limited coverage across the field.

\subsubsection{Subaru Hyper Suprime-Cam - Subaru Strategic Program (HSC-SSP) - g,r,i,z,y}

Hyper Suprime-Cam (HSC) is an optical imaging camera, with a wide field of view (1.5 deg diameter), mounted on the Subaru 8.2-m telescope at Mauna Kea, Hawaii. The HSC-Subaru Strategic Program (HSC-SSP; \citealp{2018PASJ...70S...4A}) began in 2014 and was awarded 330 nights to conduct a three-layered (wide, deep and ultra-deep) survey using a multitude of broad (\textit{grizy}) and narrowband filters. The HSC-SSP observing footprint includes four pointings in XMM-LSS, of which three are `deep', with 5$\sigma$ limiting magnitudes of 27.1, 26.5, 26.2, 25.9, 24.7 mag in \textit{grizy}, respectively. The fourth pointing is `ultradeep', collected using longer telescope integration times, with 5$\sigma$ limiting magnitudes of 27.6, 27.1, 26.9, 26.5, 25.6 mag in the same bands mentioned above. We use optical data from the \citetalias{Varadaraj_2023} catalogues who take \textit{grizy} photometry from HSC-SSP Data Release 3 \citep{Aihara_2022}. The area over XMM-LSS for which HSC optical data is available totals 4.33 deg$^2$; determined by the regions that contain overlapping NIR data (specifically from the VISTA telescope), in mind of our goal of performing a multi-wavelength study of SN Ia host galaxies. The typical seeing is $\sim 0.8$\,arcsec. The survey did not cover CDFS; therefore, the HSC catalogue for this field is derived from archival HSC data \citep{2019RNAAS...3....5N} obtained from four pointings across the field (except for HSC-\textit{R}, which constituted a single central pointing in CDFS; see \citetalias{Varadaraj_2023}, Fig. 1). The seeing is slightly poorer in this field, due to the relatively low-elevation on the sky as observed from Hawaii, and varies by $\sim 0.1$~arcsec across the field (see \citetalias{Varadaraj_2023} for details). However, as we use total magnitudes for our work, this has little effect on our results.

\subsection{Near-infrared}

\subsubsection{Visible and Infrared Survey Telescope for Astronomy (VISTA)-VIDEO - Y,J,H,K$_s$}

The VISTA Deep Extragalactic Observations (VIDEO) survey (\citealp{2013MNRAS.428.1281J}) is a near-infrared survey conducted on the VISTA telescope, a 4.1-m wide field telescope located at the Paranal Observatory, Chile. Observations were taken in the \textit{Z,Y,J,H} and $K_\mathrm{s}$ bands, with a maximum seeing of $0.9$~arcsec, and covered $\sim$ 12 deg$^2$ over three legacy fields: ES1, CDFS and XMM-LSS. For the fields relevant to this work (the latter two), VISTA pointings are split into three tiles of 1.5 deg$^2$ each (totalling $\sim$ 4.5 deg$^2$), over which our data, as for the optical, came from the \citetalias{Varadaraj_2023} catalogues. Only the \textit{Y,J,H,$K_\mathrm{s}$} bands are considered in this work, for which the average 5$\sigma$ limiting magnitudes across both fields are: $\sim$ 25.2, 24.7, 24.2 and 23.8, respectively. The exact depths for each tile in CDFS and XMM can be found in \citetalias{Varadaraj_2023}, Table 1. The reason for excluding VISTA $Z$-band data is due to incomplete survey coverage over our fields. Instead, this is supplemented by HSC-$Z$, which also probed to greater depths.

\subsection{Mid-infrared}

\subsubsection{Spitzer SERVS - 3.6, 4.5 $\mu$m}

NASA's \textit{Spitzer Space Telescope}, operational between 2003 - 2020, was an infrared observatory designed to study cool and dusty regions of the Universe that are obscured at optical wavelengths. Of the three on-board instruments, probing the shorter wavelength regime of the spectral range ($\lambda \sim 3 - 160~\mu$m), was the Infrared Array Camera (IRAC), a four-channel detector that comprised of filters centered at 3.6, 4.5, 5.8 and 8.0 $\mu$m, with an angular resolution of ${\sim}2$ arcsec. We use IRAC data\footnote{SERVS catalogues obtained from the NASA/IPAC Infrared science archive: \url{https://irsa.ipac.caltech.edu/cgi-bin/Gator/nph-scan?submit=Select&projshort=SPITZER}} from the \textit{Spitzer} Extragalactic Representative  Volume Survey (SERVS; \citealp{2012PASP..124..714M}). SERVS is a medium-deep survey that imaged 18 deg$^2$ over five legacy fields: CDFS, ES1, Elais-N1 (EN1), Lockman Hole and XMM-LSS, to a depth of $\sim$2 $\mu$Jy. SERVS was conducted during the "warm mission" phase of \textit{Spitzer's} operation, which began once the instrument ran out of liquid helium coolant. Only IRAC channels 1 (3.6 $\mu$m) and 2 (4.5 $\mu$m) continued to operate at peak performance during this period and thus, are the only two bands for which SERVS data is available. Whilst archival \textit{Spitzer} data is available at 5.8 and 8.0 $\mu$m over our studied fields, we do not include these channels in this analysis as the data are at a shallower depth and do not provide significant improvements on any host galaxy parameter constraints.

\subsubsection{Spitzer SWIRE - MIPS 24 $\mu$m}

The \textit{Spitzer} Wide-area InfraRed Extragalactic (SWIRE; \citealp{2003PASP..115..897L}) survey is one of the largest programs undertaken during the early stages of the telescopes operation. Designed with the aim of observing galaxies to an extended area and depth, SWIRE imaged $\sim$49 deg$^2$ across six legacy fields: CDFS, ES1, EN1, ELAIS-N2, Lockman Hole, XMM-LSS. SWIRE data was collected using IRAC and the Multiband Imaging Photometer (MIPS) that operated at wavelengths of 24, 70 and 160~$\mu$m. The angular resolutions for each of these wavebands are $\sim 6$, 18 and 40~arcsec, respectively. In this work, we only use MIPS 24 $\mu$m data obtained from the Herschel Extragalactic Legacy Project (HELP; \citealp{2021MNRAS.507..129S}) database\footnote{\url{https://herschel.sussex.ac.uk} \label{help}}. The 70 $\mu$m channel is excluded due to its poor sensitivity and incomplete survey coverage in our fields. At 160 $\mu$m, we use deeper \textit{Herschel} observations in the corresponding band.

\subsection{Far-infrared}

\subsubsection{Herschel PACS - 100, 160 $\mu$m}

The \textit{Herschel Space Observatory}, built by the European Space Agency and equipped with a 3.5-m diameter mirror, was the largest infrared mission of its generation ever launched. Mapping the cosmos from the far-infrared to sub-mm domain, \textit{Herschel} was designed with the intention of studying cold and dusty regions of space, from stellar nurseries to the evolution of objects on galactic scales. One of the on-board instruments was the Photoconductor Array Camera and Spectrometer (PACS) instrument -- an imaging camera with filters centered at 70, 100 and 160 $\mu$m. The typical full width at half maximum (FWHM) for each of these wavebands are 5.5, 6.7 and 11~arcsec, respectively. In mind of our goal of probing the dust emission curve of galaxy spectra, we include PACS 100 and 160 $\mu$m data from the HELP\footref{help} database. As such, our final catalogues are derived from a combination \textit{Herschel} surveys, namely, the Herschel Multi-tiered Extragalactic Survey \citep[HERMES; ][]{HERMES} and the PACS Evolutionary Probe \citep[PEP; ][]{PEP}. The previously mentioned reasons for excluding \textit{Spitzer} 70 micron also apply here.

\subsubsection{Herschel SPIRE - 250, 350, 500 $\mu$m} 

To sample the full dust emission curve, complementary to PACS, the \textit{Herschel} Spectral and Photometric Imaging REceiver (SPIRE) instrument covered the 250, 350 and 500 $\mu$m spectral bands.
The angular resolutions 
for each band are 18.2, 24.9 and 36.3~arcsec, respectively. All channels are included in our analysis and are, as above, obtained from HELP catalogues, derived from HERMES. To summarise, all photometric wavebands and the part of a galaxy spectral energy distribution (SED) they probe are illustrated in Fig. \ref{frank}.

\section{Photometry and final catalogue compilation} \label{phot}

\subsection{\textit{Herschel} and \textit{Spitzer} photometric measurements}

In this section, we detail the methods used to obtain photometry from the mid- to far-infrared data for objects that have non-detections at these wavelengths. The catalogues outlined in Section~\ref{hostcat} only include host galaxy photometry for objects `detected' by \textit{Herschel} and \textit{Spitzer}. Therefore, including and performing SED fitting only on galaxies that only have significant detections in these bands introduces a bias toward studying dusty (likely star-forming) environments. To prevent this bias, and mitigate against degeneracies encountered when using optical data alone, we measure photometry for objects that are not detected at the longer wavelengths and use these fluxes (in conjunction with the catalogues) to help constrain the dust emission.

For the \textit{Spitzer} IRAC channels 1 and 2, we use \texttt{SExtractor} \citep{1996A&AS..117..393B}, in dual-image mode, using 3.6 $\mu$m as the detection image and measuring from the 4.5 $\mu$m filter. Running \texttt{SExtractor} in dual-image mode reduces the chance of there being aperture mismatches between galaxies as the same \texttt{FLUX$\_$AUTO} aperture is used for any given source imaged across the different wavebands. In addition, a run of \texttt{SExtractor} in single-image mode (i.e. on IRAC channels 1 and 2 separately) results in a handful of objects ($\sim$ 6$\%$ of the total sample) detected at 3.6 $\mu$m but not at 4.5 $\mu$m; as expected for hosts situated in the lower luminosity and/or higher redshift region of the parameter space. Dual-image mode avoids this, ensuring a more complete and unbiased sample that includes low-signal-to-noise measurements. The parameter values in the input file for \texttt{SExtractor} are chosen following \cite{Lacy_2005} and \cite{2012PASP..124..714M} and are listed in Appendix \ref{A}.

At wavelengths $\lambda$ $\gtrsim$ 8 $\mu$m, the flux is spread over a much larger area due to the poorer resolution. Therefore, to extract fluxes in the \textit{Spitzer} MIPS 24 $\mu$m band, 7~arcsec diameter apertures are used, with aperture corrections applied following guidelines found in the MIPS instrument handbook (\citealp{mips_handbook}\footnote{The MIPS instrument handbook can be found here: \url{https://irsa.ipac.caltech.edu/data/SPITZER/docs/mips/mipsinstrumenthandbook/MIPS_Instrument_Handbook.pdf} \label{mipshandbook}}). For PACS 100~$\mu$m and 160 $\mu$m, aperture sizes are chosen to be similar to the point spread function (PSF) FWHM, with diameters  of 7~arcsec and 12~arcsec, respectively. The exact aperture corrections are derived using the PSF image for each band\footref{help}, assuming the galaxies are unresolved.

Far-infrared \textit{Herschel} photometry at 250, 350 and 500 $\mu$m is performed on SPIRE point source maps, which are calibrated in units of Jy/beam. 
We extract fluxes by measuring the pixel value at the coordinates of the DES-SN host galaxies, which represent the peak flux density of a point source positioned at the centre of that pixel. The corresponding errors are derived via a two step process. First, a background distribution is determined by measuring the values in the nearest 300 pixels (an arbitrary choice) to each source position. Due to the problem of source confusion at these wavelengths, from the presence and blending of multiple unresolved galaxies within a single SPIRE beam, the background histogram resembles a skewed normal distribution; with an excess of positive source counts. Therefore, the second step in determining the error on the SPIRE fluxes is to fit the distribution with an asymmetric Gaussian. We conservatively take the right-hand side (i.e. larger) standard deviation as our error term, to account for the issue of source confusion.

\subsection{Catalogue creation}

The final catalogues for each field are assembled via crossmatching each of the aforementioned multi-wavelength galaxy datasets to the DES-SN host coordinates (from the \citetalias{2020MNRAS.495.4040W} catalogue). Optical and near- to mid-infrared data are positionally crossmatched to DES using a 1~arcsec matching radius. Following \cite{Davies_2021}, data for wavelengths at and above MIPS 24 $\mu$m are matched with a 5~arcsec radius, using nearest neighbour association to account for the poorer positional accuracy.

Having compiled the catalogues, next we correct the fluxes in each filter for dust extinction due to the Milky Way. The reddening, $E(B-V)$, is calculated for each DES galaxy sightline using the dust maps presented in \cite{2011ApJ...737..103S}. The extinction in each photometric band is then estimated and corrected for using the \cite{1999PASP..111...63F} dust law, with $R_V = 3.1$, the Milky Way average. As a last step, we impose a 5$\%$ error floor for measurement uncertainties that fall below this threshold. This is to account for the fact that (i) uncertainties in the zero-point for broad-band imaging, which is dependent on the colour of objects; and (ii) when performing SED fitting, the template set does not contain a perfect model for the SEDs of real galaxies.  The final sample of SN host galaxies for which data is available at each wavelength is summarised in Table \ref{numbers}. We note that the variations in SN host numbers with data in each band is due to different area coverage of the DES fields by the various telescopes and not due to sensitivity differences. Of this sample, a total of 501 hosts have data at \textit{every} wavelength and
we designate these galaxies our `gold' sample.

\begin{table*}
	\centering
	\caption{Number of DES-SN galaxies for which data is available (in at least a single band) across the different surveys. These numbers are indicative of host galaxy data that comes from a combination of catalogue data and photometric measurements.}
	\label{number}
	\begin{tabular}{p{2.5cm} p{6cm} p{2.5cm} p{2.5cm}}  
		\hline
		\hline
        &  & \multicolumn{2}{c}{\textbf{Field (total $\#$ of DES-SN host galaxies\textsuperscript{\textdagger})}} \\[1ex] \cline{3-4}\\[0.001ex] 
        &  & \multicolumn{1}{c}{CDFS (271)} & \multicolumn{1}{c}{XMM (230)} \\
        \hline
          & Survey/Wavelength & \\
		\hline
        \hline
		Optical & CFHT-LS, $u^\star$ & \multicolumn{1}{c}{--} &
        \multicolumn{1}{c}{263} \\
        & HSC-SSP, \textit{grizy} & \multicolumn{1}{c}{274} & \multicolumn{1}{c}{269} \\
        & VST VOICE, \textit{ugri} & \multicolumn{1}{c}{280} & \multicolumn{1}{c}{--} \\
        \hline
		Near-infrared & VISTA VIDEO, \textit{YJHK$_\mathrm{s}$} & \multicolumn{1}{c}{280} & \multicolumn{1}{c}{269} \\
        \hline
		 Mid-infrared & \raggedright \textit{Spitzer}  SERVS, \textit{3.6,4.5}$\mu m$ & \multicolumn{1}{c}{339} & \multicolumn{1}{c}{319} \\
		 & \raggedright \textit{Spitzer} SWIRE, MIPS \textit{24}$\mu m$ & \multicolumn{1}{c}{445} & \multicolumn{1}{c}{369} \\
        \hline
		Far-infrared & \raggedright \textit{Herschel} PACS, \textit{100,160}$\mu m$ & \multicolumn{1}{c}{453} & \multicolumn{1}{c}{521} \\
        & \raggedright \textit{Herschel} SPIRE, \textit{250,350,500}$\mu m$ & \multicolumn{1}{c}{534} & \multicolumn{1}{c}{523}\\
		\hline
	\end{tabular}
    \parbox{\linewidth}{\small \textsuperscript{\textdagger} In brackets next to each respective field are the total number of DES SN Ia host galaxies that have data for across \textit{all} bands and thus, form the final `gold' sample for this analysis i.e. a subtotal of 501 galaxies.}
\end{table*} \label{numbers}

\section{Galaxy Spectral Modelling with \textsc{bagpipes}} \label{bag}

\textsc{Bagpipes} (Bayesian Analysis of Galaxies for Physical Inference and Parameter EStimation; \citealp{Carnall_2018}) is a galaxy SED fitting Python code that uses Bayesian techniques to fit models to photometric and spectroscopic data spanning from the ultraviolet to microwave regime. At the foundational level, \textsc{Bagpipes} constructs galaxy SEDs using the updated 2016 \cite{2003MNRAS.344.1000B} stellar population synthesis models, with a \cite{2001ASPC..228..187K} initial mass function. In addition, there are user-specified parameters pertaining to components such as star formation histories (SFH), nebular emission and dust attenuation that can be used to fine-tune the complexity of the model. We use {\sc Bagpipes} with the nested-sampling algorithm \textsc{MultiNest} with a configuration of 1000 live points. The outputs are checked to ensure the posteriors are reasonable and the sampling has converged. An example of a galaxy SED fit and posterior distribution output from \textsc{Bagpipes} is presented in Appendix \ref{appendix:B}. 

We assume a log-normal SFH, as also used in \cite{Meldorf_2022}, which requires a smooth rise and fall of star formation over cosmic history, as opposed to a more rapidly quenching system. This choice is made following \cite{Gladders_2013} who, based on the observation that the global SFH of \textit{all} galaxies follows a shape evolution in time that is well fit by a log-normal distribution, suggest the same may apply to individual sources. \cite{Diemer_2017} investigate this further and find that this parametric form produces successful fits to cosmological simulations. As we are primarily interested in the constraints on galaxy dust parameters that are output via SED fitting, we restrict the SFH (and all other relevant inputs) to a single functional form throughout all \textsc{Bagpipes} runs; the priors on which are summarised in Table \ref{bagpipes}. The  redshift of each galaxy is fixed to the spectroscopic host redshift from the DES-SN5YR data release\footref{desdr}.

Regarding dust extinction and/or attenuation curves, they are typically expressed in terms of two parameters: (i) $A_\lambda$, the total attenuation at a reference wavelength, $\lambda$, which acts as the normalisation of the curve. It is common practice to take the attenuation in the $V-$band, $A_V$ ($\lambda \sim 5500$\r{A}). (ii) a parameter that characterises the slope of the dust law, usually represented by the total-to-selective ratio, $R_V$. The value of $R_V$ is dependent upon the size of the dust particles. Smaller values of $R_V$ correspond to steeper dust curves. This indicates the presence of smaller dust grains, resulting in a more wavelength-dependent extinction of light, i.e. greater reddening of light due to dust. In contrast, larger values of $R_V$, which yield shallower dust curves, coincide with larger dust grains and result in light being extinguished with a lower dependence on wavelength.

The analysis presented here uses the \citealp{2000ApJ...539..718C} (hereafter \citetalias{2000ApJ...539..718C}) dust attenuation law with a variable slope. The \citetalias{2000ApJ...539..718C} model adopts a two-component form, whereby different amounts of attenuation are applied on account of stellar population age. Young ($\lesssim$ 10 Myr) stars are doubly attenuated, both by dust in their natal clouds and the interstellar medium, whereas older stellar populations are only attenuated by the latter. Mathematically, the \citetalias{2000ApJ...539..718C} dust law is defined in \textsc{Bagpipes} as,

\begin{equation} \label{eq:cf00}
\frac{k(\lambda)}{R_V} = (\lambda/5500\text{\r{A}})^{-n} ,
\end{equation}

where $n$ is the slope of the dust law. Our priors on $n$ (Table \ref{bagpipes}) are chosen by comparing the different values of $n$ to known values of $R_V$ found from studies of the Milky Way. 

In addition to our fiducial \textsc{Bagpipes} run using the \citetalias{2000ApJ...539..718C} dust model, we test several other parametric dust models currently supported in \textsc{Bagpipes}. The first is the Cardelli (\citealp{1989ApJ...345..245C}) Milky Way dust law, modelled from studies of interstellar extinction in the local Universe. This is the simplest dust law in the sense that it only accounts for the effects of dust absorption and scattering of light. The second \cite{2000ApJ...533..682C} dust law implementation introduces a further degree of complexity, such that it accounts for the effects of dust attenuation i.e., the scattering of light back into the line of sight and star-dust geometry. The Calzetti law is commonly applied in the analyses of star-forming galaxies.

The \cite{1989ApJ...345..245C} and \cite{2000ApJ...533..682C} dust laws are both often characterised by a single degree of freedom, allowing the normalisation $A_V$ to vary but keeping the slope of the dust law fixed (with values in the literature typically assumed to be $R_V = 3.10$ and $R_V = 4.05$, respectively). There is evidence however that the slope of the dust attenuation curve varies across different galaxies as well as for different lines of sight within a single galaxy \citep{2013ApJ...776....7G,2016ApJ...821...78S,2017ApJ...847..102Y,2018ApJ...859...11S}. There are dust models implemented in \textsc{Bagpipes} (in addition to \citetalias{2000ApJ...539..718C}) that account for this variation and allow for greater flexibility in fitting. The \cite{2018ApJ...859...11S} dust law, for example, is a modification of the \cite{2000ApJ...533..682C} curve in two aspects. The first allows the slope to deviate from the \cite{2000ApJ...533..682C} model through the introduction of a power-law term with exponent $\delta$ (note, setting $\delta$ = 0 reproduces the \cite{2000ApJ...533..682C} dust parameterisation). A second modification introduces a ultraviolet (UV) "bump" centered at 2175 \r{A}. 

\begin{table*} 
	\centering
	\caption{Input global parameters and priors for \textsc{bagpipes} SED modelling. All priors are uniform unless explicitly stated otherwise. Dust model component-specific parameters are associated via colour coding.}
	\label{bagpipes}
	\begin{tabular}{p{2.4cm} p{2.5cm} p{9.1cm} p{1.5cm}}  
		\hline
		\hline
		\textbf{Component} & \textbf{Symbol} & \textbf{Description} & \textbf{Prior} \\
		\hline
		Global & $z$ & Redshift & \multicolumn{1}{c}{$z_{\mathrm{spec}}$}\\
        \hline
		\multirow[t]{4}{2.2cm}{Star-formation history; lognormal} & t$_{\mathrm{max}}$ (Gyr) & \raggedright Age of Universe at peak star-formation & \multicolumn{1}{c}{(0.1, 15)}\\
		& FWHM (Gyr) & \raggedright Full width at half maximum of star-formation & \multicolumn{1}{c}{(0.1, 20)}\\
		 & log$_{10}(M_\star/M_\odot)$ & Total stellar mass formed & \multicolumn{1}{c}{(1, 15)}\\
		 & $Z_\star/Z_\odot$ & Stellar metallicity & \multicolumn{1}{c}{(0, 3)}\\
        \hline
		Nebular & log$_{10}U$ & Ionization parameter & \multicolumn{1}{c}{(-4, -2)}\\
        \hline
		Dust & Type & Cardelli; Calzetti; \textcolor{Mahogany}{Charlot $\&$ Fall}; \textcolor{MidnightBlue}{Salim} &  \multicolumn{1}{c}{--}\\
        & $A_V$ (mag) & \raggedright Absolute $V$-band attenuation & \multicolumn{1}{c}{(0, 6)} \\
        & $\epsilon$ & \raggedright Multiplicative constant on $A_V$ for stars in birth clouds (t$_{\mathrm{BC}} < 10$ Myr)& \multicolumn{1}{c}{1.0}\\
        & $\mathcal{Q}_{\mathrm{PAH}}$ & \raggedright Polycyclic aromatic hydrocarbon (PAH) mass fraction & \multicolumn{1}{c}{(0.1, 4.58)}\\
        & $\mathcal{U}_{\mathrm{min}}$ & \raggedright Minimum starlight intensity that dust is exposed to & \multicolumn{1}{c}{(0.1, 25.)}\\
        & $\gamma_{\mathrm{e}}$ &  Fraction of starlight at $\mathcal{U}_{\mathrm{min}}$ & \multicolumn{1}{c}{(0.0005, 1.)}\\
        & \textcolor{Mahogany}{\textit{n}} & Slope of the dust attenuation law &\multicolumn{1}{c}{(0.3, 1.2)}\\
        & \textcolor{MidnightBlue}{$\delta$} & \raggedright Deviation of the slope of the dust law from the Calzetti law & \multicolumn{1}{c}{(-0.7, 0.3)}\\
        & \textcolor{MidnightBlue}{\textit{B}}& 2175\r{A} bump strength & \multicolumn{1}{c}{(0, 5)} \\
		\hline
	\end{tabular}
\end{table*}

\section{Results I: Host galaxy parameters determined with \textit{Herschel} and \textit{Spitzer}} \label{results}

In this section we investigate the impact that adding \textit{Herschel} and \textit{Spitzer} data has on a selection of host galaxy parameters.

\subsection{Dust parameter constraints}

\subsubsection{Dust attenuation, $A_V$}

\begin{figure*}
  \begin{subfigure}[b]{\linewidth}
    \centering
    \includegraphics[width=0.7\linewidth]{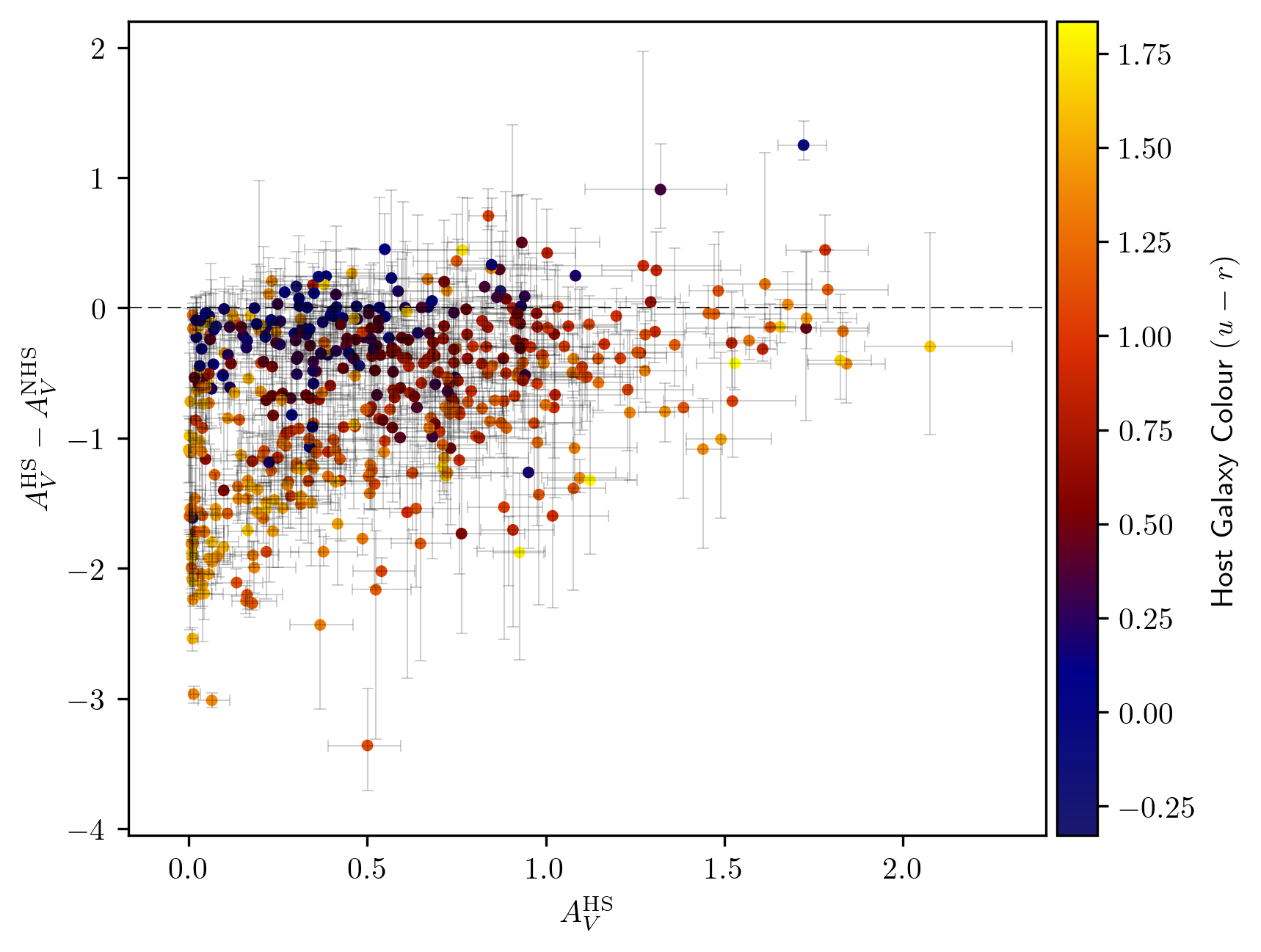} 
    \label{Avres} 
  \end{subfigure}
  \vspace{1em} 
  \begin{subfigure}[b]{\linewidth}
    \centering
    \includegraphics[width=0.7\linewidth]{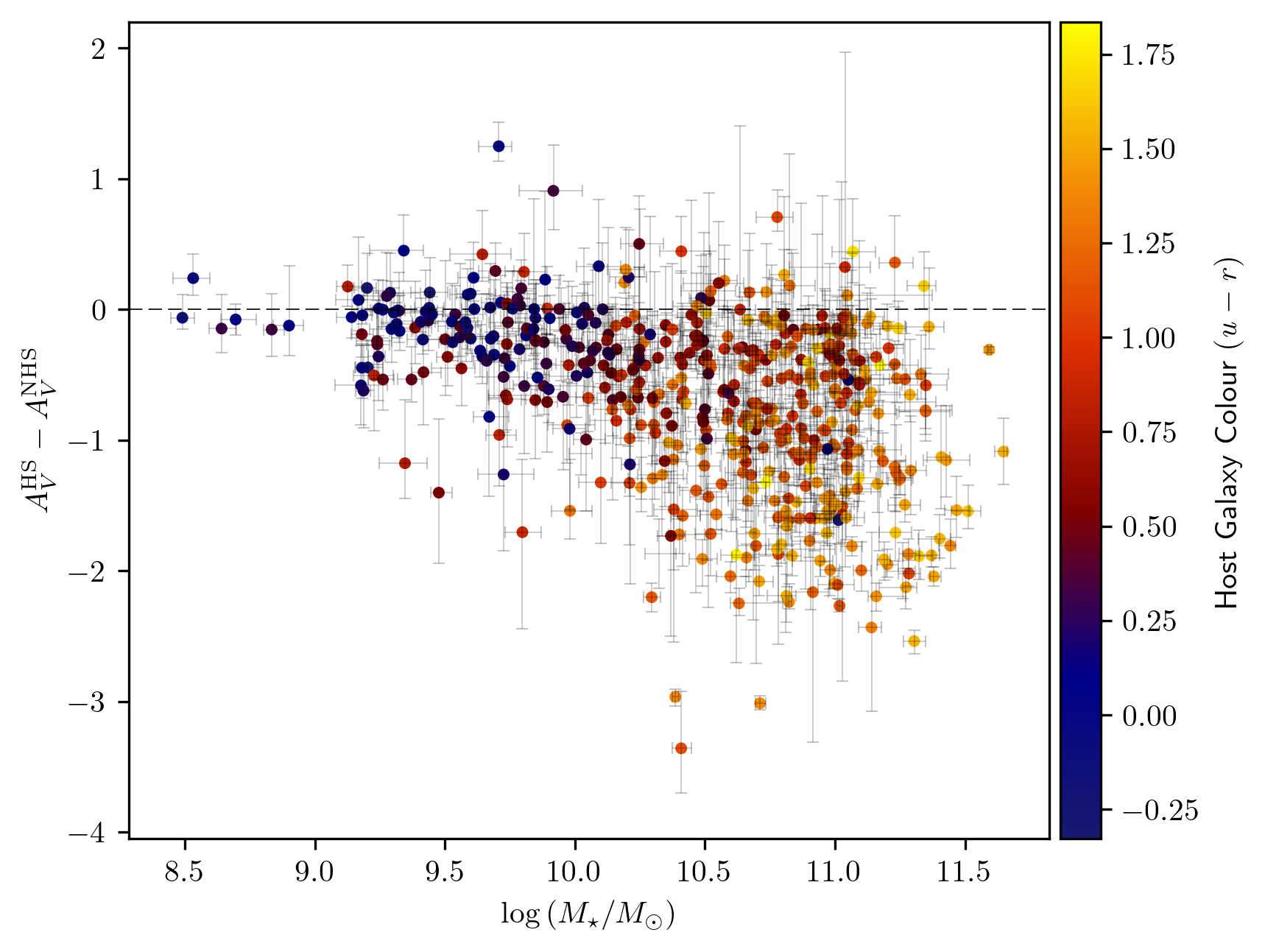} 
    \label{massres} 
  \end{subfigure}
  \caption{Residuals in $A_V$ showing the effect of including and excluding \textit{Herschel} and \textit{Spitzer} data as a function of $A_V$ determined using \textit{Herschel} and \textit{Spitzer} (top) and stellar mass (bottom). `HS' refers to "\textit{Herschel} and \textit{Spitzer}" and refers to when all wavebands are used to obtain host galaxy parameter constraints. `NHS' refers to "No \textit{Herschel} and \textit{Spitzer}" for the cases when far-infrared data is excluded. If neither of these two superscripts are used (such is the case for the x-axis label on the bottom plot in this Figure), this implies \textbf{all} wavebands are used in obtaining the relevant host galaxy constraint. The samples are colour-coded as a function of host galaxy rest-frame \textit{u -- r} colour. We note that the axis is truncated to omit a single data point that lies at $A_V^{\mathrm{HS}} > 3$ in order to better highlight the main result displayed in this Figure.} 
  \label{res_dust} 
\end{figure*}

We estimate the best-fit host galaxy dust attenuation values obtained when including \textit{Herschel} and \textit{Spitzer} data ($A_V^{\mathrm{HS}}$; `HS' stands for `\textit{Herschel} and \textit{Spitzer}') and excluding it ($A_V^{\mathrm{NHS}}$; `NHS' stands for `No \textit{Herschel} and \textit{Spitzer}') and compare the two measurements. The comparisons are shown in Fig.~\ref{res_dust} as a function of both $A_V^{\mathrm{HS}}$ (which serves as a more reliable proxy for the intrinsic host galaxy $A_V$ as it is derived using all available photometric wavebands) and stellar mass. For the residuals as a function of $A_V^{\mathrm{HS}}$ (top), the largest discrepancies occur for objects that have low attenuation when using the long-wavelength data ($A_V^{\mathrm{HS}} \lesssim 0.4$ mag). Colour-coding the data as a function of measured host galaxy rest-frame $u-r$ colour reveals that objects in this low-$A_V^{\mathrm{HS}}$ regime are intrinsically the reddest host galaxies ($u-r > 1$). This observed trend arises as a consequence of the degeneracy between age, metallicity and dust, as all three factors can act to redden the colour of a galaxy. Therefore, when performing SED fitting on intrinsically red galaxies (i.e. predominantly redder not because of dust but likely because of the presence of an older stellar population or metallicity effects), the $A_V$ values can be overestimated in the absence of far-infrared data, in order for the fitted SED model to best reproduce the observed galaxy colours. $A_V$ discrepancies decrease for high $A_V^{\mathrm{HS}}$ galaxies as, despite still requiring higher dust attenuation when excluding \textit{Herschel} and \textit{Spitzer}, the red galaxy colours are now predominantly due to dust, so predictions here are more likely to coincide with the intrinsic host $A_V$.

For the same reasons outlined above, the largest $A_V$ residual differences are observed for high mass host galaxies ($\mathrm{log}_{10}(M_\star/M_\odot) > 10$), as seen in the bottom panel of Fig.~\ref{res_dust}. Massive galaxies are comprised of primarily older stellar populations, which emit most of their light at redder wavelengths. Therefore, the values of $A_V$ for more massive, redder hosts can be overestimated as the absence of far-infrared data makes it challenging to break the age-metallicity-dust degeneracy.

\subsubsection{Dust law slope}

To check that the results of our analysis are not affected by the choice of dust law, we replace our fiducial choice of the \citetalias{2000ApJ...539..718C} law with all other parametric forms implemented in \textsc{Bagpipes}. We highlight that the main results of our analysis are not affected by the choice of dust law. A second reason that motivates this is that it informs us (when including \textit{Herschel} and \textit{Spitzer}) how well we are able to constrain parameters characterising the slope of the dust attenuation law. To do so, we perform a Bayesian model comparison of the different dust models available in \textsc{Bagpipes}, \textit{with} and \textit{without} the additional dust-slope free parameters (as well as Cardelli vs. Calzetti for which the slope is fixed for completeness). Using the Jeffrey's scale on the Bayesian evidences for each galaxy and dust law choice, strong preference for any given dust law is defined as log$_{10}$(B$_{1,2}$) > 5 (and its reciprocal), where B$_{1,2}$ is the Bayes factor. We present the results in Table. \ref{tab:dust_law_comparison}. Across all dust models, we find that our data show no preference for an additional dust-slope-free parameter and that $R_V$, $n$ and $\delta$ remain largely unconstrained, even when utilising the extensive multi-wavelength dataset that we have. This finding is supported by the fact that when running the same dust law but changing whether the slope of the dust law is free or fixed, most galaxies in our sample show no preference for either choice. Invoking Occam's Razor, the model with the fewest number of free parameters should be preferred i.e., dust models that fix the slope of the dust law. We therefore urge caution when fitting more free parameters for the dust attenuation models than the data can justifiably constrain.

\begin{table}
\centering
\resizebox{1.03\columnwidth}{!}{
\begin{tabular}{lccc}
\hline
\hline
\textbf{Comparison} & \textbf{Dust Law 1 Preferred} & \textbf{Dust Law 2 Preferred} & \textbf{No Preference} \\
\hline
Cardelli vs Calzetti & 35 (Car) & 18 (Cal) & 448 \\
CF00 fixed $n$ vs free $n$ & 1 (fixed) & 4 (free) & 496 \\
Salim fixed $\delta$ vs free $\delta$ & 2 (fixed) & 52 (free) & 447 \\
\hline
\end{tabular}
}
\caption{Number of objects preferring each dust law or showing no preference. These numbers are determined using Jeffrey's scale on the Bayesian evidences for each galaxy and dust law choice. Strong preference for any given dust law is defined as log$_{10}$(B$_{1,2}$) > 5 (and its reciprocal), where B$_{1,2}$ is the Bayes factor.}
\label{tab:dust_law_comparison}
\end{table}

\subsection{Stellar mass constraints}

\begin{figure}
    \centering
    \includegraphics[width = 82mm]{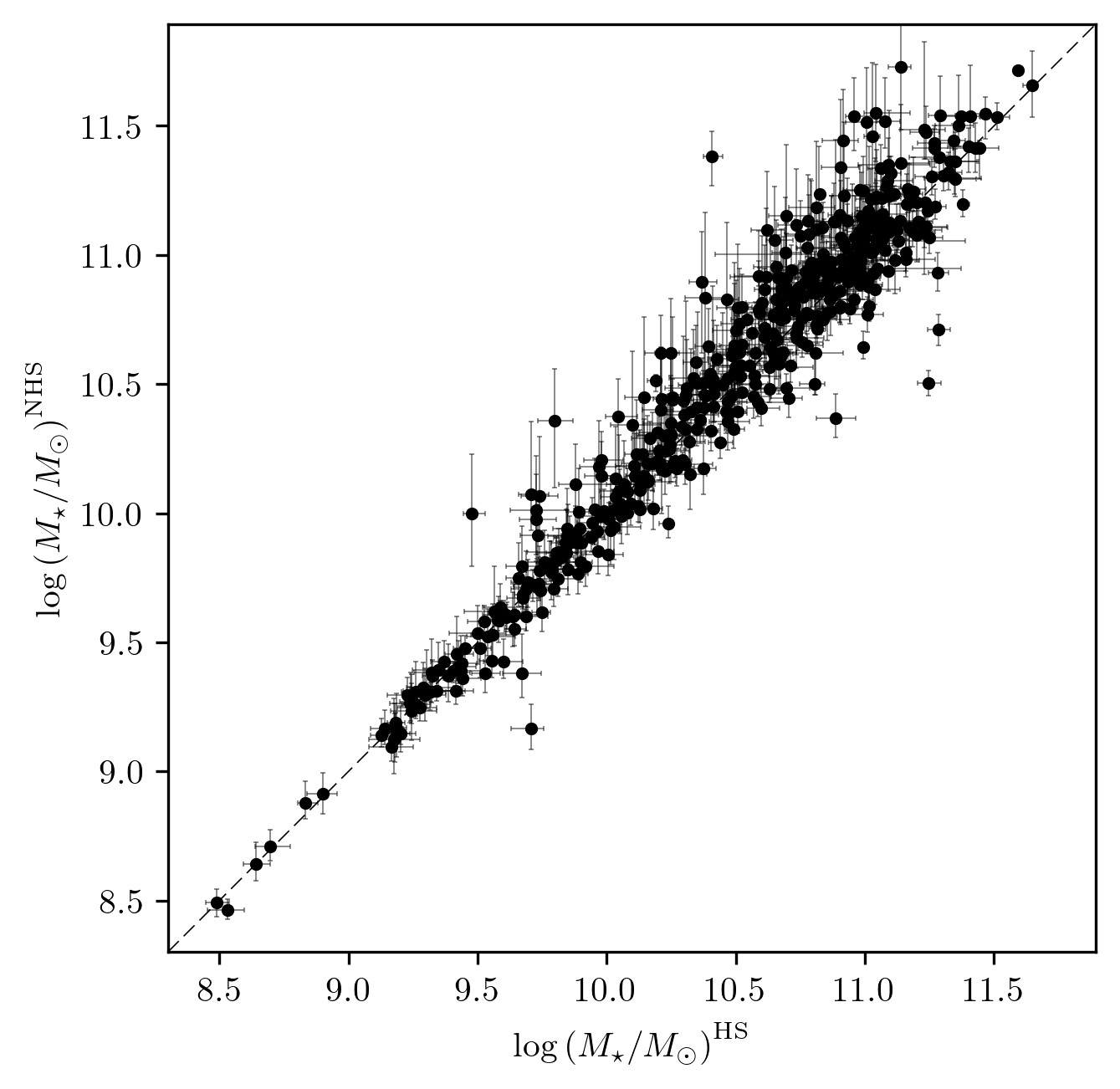}
    \caption{Comparison of host galaxy stellar masses constrained when including (`HS') and excluding (`NHS') \textit{Herschel} and \textit{Spitzer} data.}
    \label{stellarmass}
\end{figure}

In Fig.~\ref{stellarmass}, we show a comparison of stellar mass estimates, measured using the long-wavelength data (log($M_\star/M_\odot$)$^{\mathrm{HS}}$) versus when we exclude it (log($M_\star/M_\odot$)$^{\mathrm{NHS}}$) 
. We find that, unlike for dust, stellar mass values derived using just the optical/NIR (\textit{ugrizy}YJHK$_\mathrm{s}$) are consistent with those derived when including longer wavebands. Stellar mass is strongly correlated with the NIR because the bulk of a galaxy's mass is comprised of older (redder) stellar populations that emit primarily at these wavelengths; where the effects of dust are also minimal (particularly filters probing the far end of the NIR which, in our sample, is VISTA K$_\mathrm{s}$). The scatter increases toward the high stellar mass end, with the stellar masses being higher when excluding \textit{Herschel} and \textit{Spitzer} data. We find this happens for a combination of reasons. First, a greater proportion of high-mass objects are at higher redshifts ($z > 0.5$). An increase in redshift means that the light emitted from the stars responsible for the bulk of the stellar mass begins to be redshifted out of the  near-infrared filters, resulting in less robust mass estimates. Whereas, the inclusion of \textit{Spitzer} data mitigates this effect. Secondly, without far-IR data, the amount of galactic dust attenuation tends to be overestimated (Fig. \ref{res_dust}). This means that there are greater degrees of freedom in the SED model posterior for higher stellar masses, as a more dust obscured galaxy can imply a more massive galaxy, just with a greater fraction of extinguished light.

\subsection{Star formation rate constraints}

\begin{figure}
    \centering
    \includegraphics[width = 90mm]{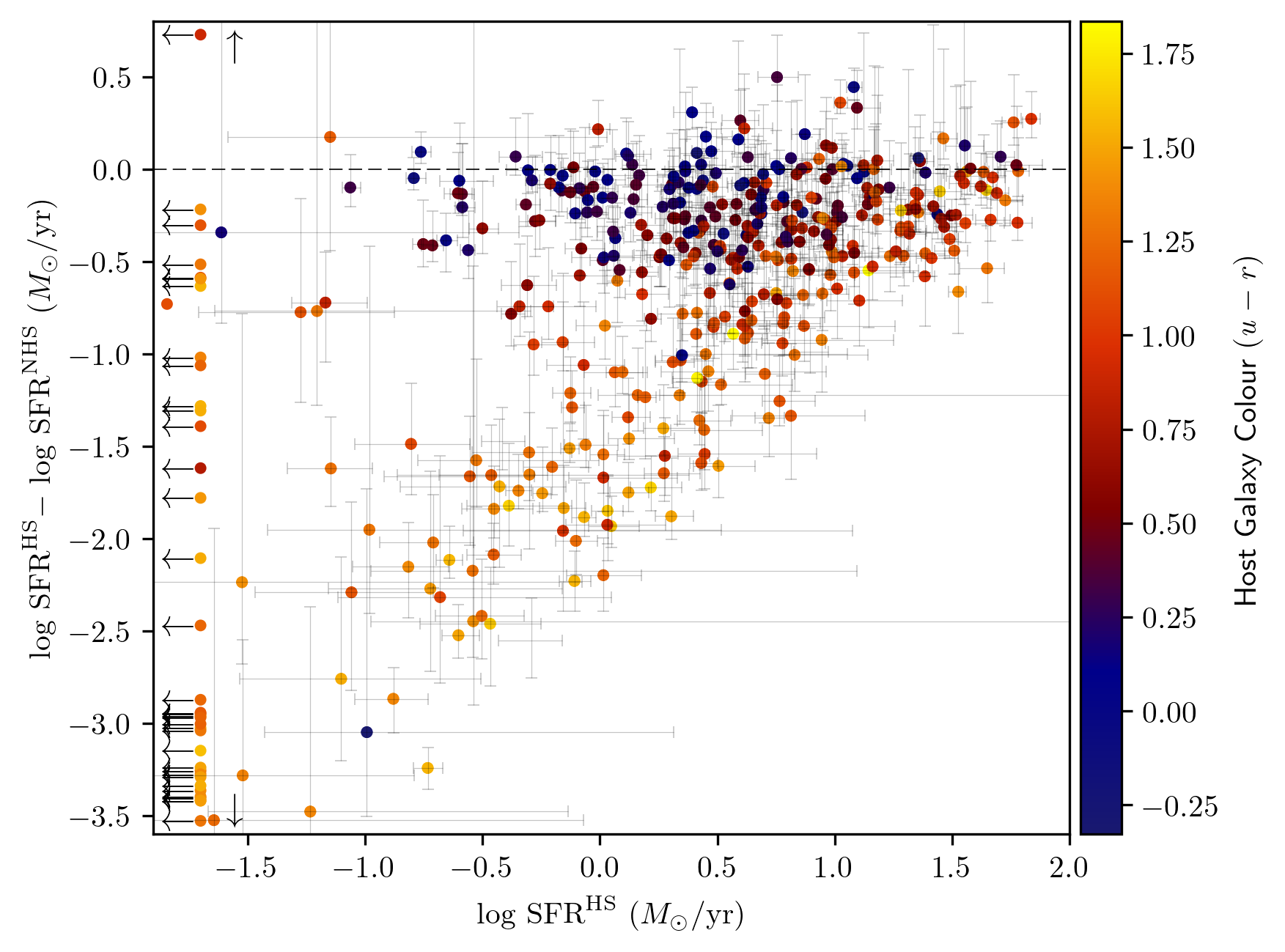}
    \caption{Residuals in log(SFR) showing the effect of including and excluding \textit{Herschel} and \textit{Spitzer} data as a function of log(SFR$^{\mathrm{HS}}$). To reiterate, `HS' refers to "\textit{Herschel} and \textit{Spitzer}" and is used when all wavebands are used to obtain host galaxy parameter constraints. `NHS' refers to "No \textit{Herschel} and \textit{Spitzer}" for the cases when far-infrared data is excluded. The samples are colour-coded as a function of host galaxy \textit{u -- r} colour. Objects with log(SFR$^{\mathrm{HS}}$) < $-$1.7 are set to limits at this value. The axis is truncated for visualisation purposes to remove objects with residuals $\Delta$log(SFR) > 0.8 and $\Delta$log(SFR) < $-$3.6.}
    \label{sfr}
\end{figure}

The residuals in star formation rate (SFR) obtained with (SFR$^\mathrm{HS}$) and without (SFR$^\mathrm{NHS}$) \textit{Herschel} and \textit{Spitzer}, are shown in Fig.~\ref{sfr} as a function of SFR$^\mathrm{HS}$. The largest SFR differences occur for objects that have little ongoing star formation, SFR$^\mathrm{HS} \lesssim 10$ M$_\odot/\mathrm{yr}$, indicative of passive galaxies that have perhaps undergone quenching. Galaxies in this part of the parameter space are also the reddest ($u - r > 1$). This observed trend is again a subsequent effect of the age-metallicity-dust degeneracy, where passive galaxies are mistaken for  dust-obscured, actively star-forming galaxies, thus leading to an over-prediction of SFR (and, in few cases, vice versa).

\section{Results II: Correlations between SN Ia colour and host galaxy dust} \label{results2}

In the context of reconciling SN Ia properties with their host galaxy environments, we investigate the link between SN colour, $c$ and host galaxy dust. In addition to global dust attenuation, $A_V^{\mathrm{HS}}$, we introduce the quantity `specific dust mass'. This is defined as the galaxy dust mass, $M_{\mathrm{dust}}$, normalised by total stellar mass, $M_\star$. We estimate the galaxy dust mass, $M_{\mathrm{dust}}$, as 

\begin{equation}
    M_{\mathrm{dust}} = \frac{F_\nu D_{\rm L}^2}{\kappa_\nu B_\nu (T_\mathrm{d})} \equiv \frac{{L}_\nu}{4\pi\kappa_\nu B_\nu (T_\mathrm{d})},
\end{equation}

where $F_\nu$ ($L_\nu$) is the flux (luminosity) density at frequency, $\nu$, measured from the best fit SED models using {\sc Bagpipes},  $D_{\rm L}$ is the luminosity distance of the source, $\kappa_\nu$ is the dust mass opacity coefficient and $B_\nu$ is the Planck function. We take $\nu = 3.53 \times 10^{11}$ Hz (850 $\mu$m) and characteristic dust temperature, $T_\mathrm{d} = 30$ K. $\kappa_\nu$ is set to 0.077 m$^2$ kg$^{-1}$ for consistency with \citealp{2000MNRAS.315..115D, 2001MNRAS.327..697D}, who choose this value as it is intermediate to values found for graphites and silicates (\citealp{1984ApJ...285...89D, 1993MNRAS.263..607H}). 

\begin{figure*} 
  \begin{subfigure}[b]{.49\linewidth}
    \centering
    \includegraphics[width=1.05\linewidth]{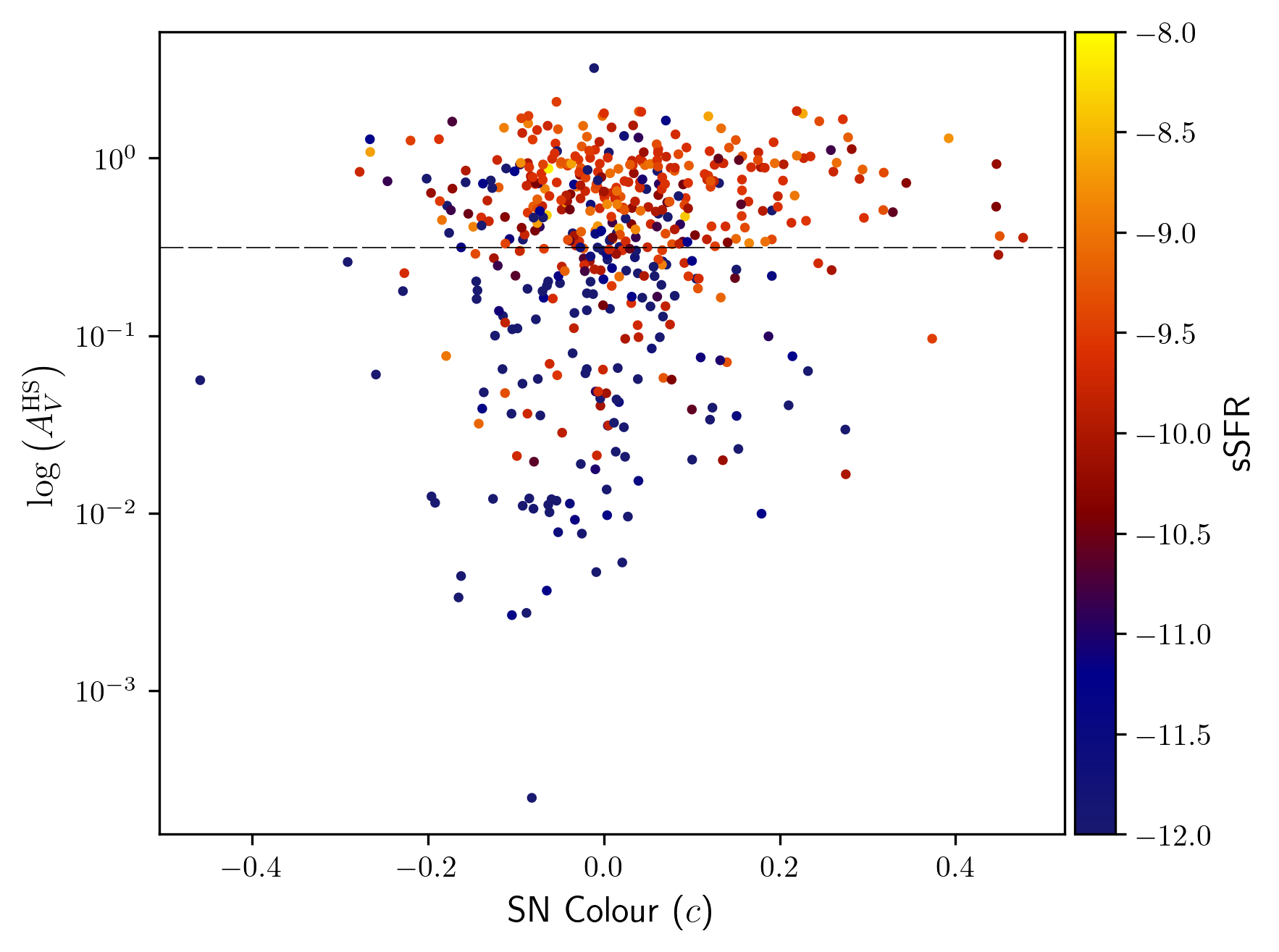} 
  \end{subfigure} \hfill
  \begin{subfigure}[b]{.48\linewidth}
    \centering
    \includegraphics[width=1.05\linewidth]{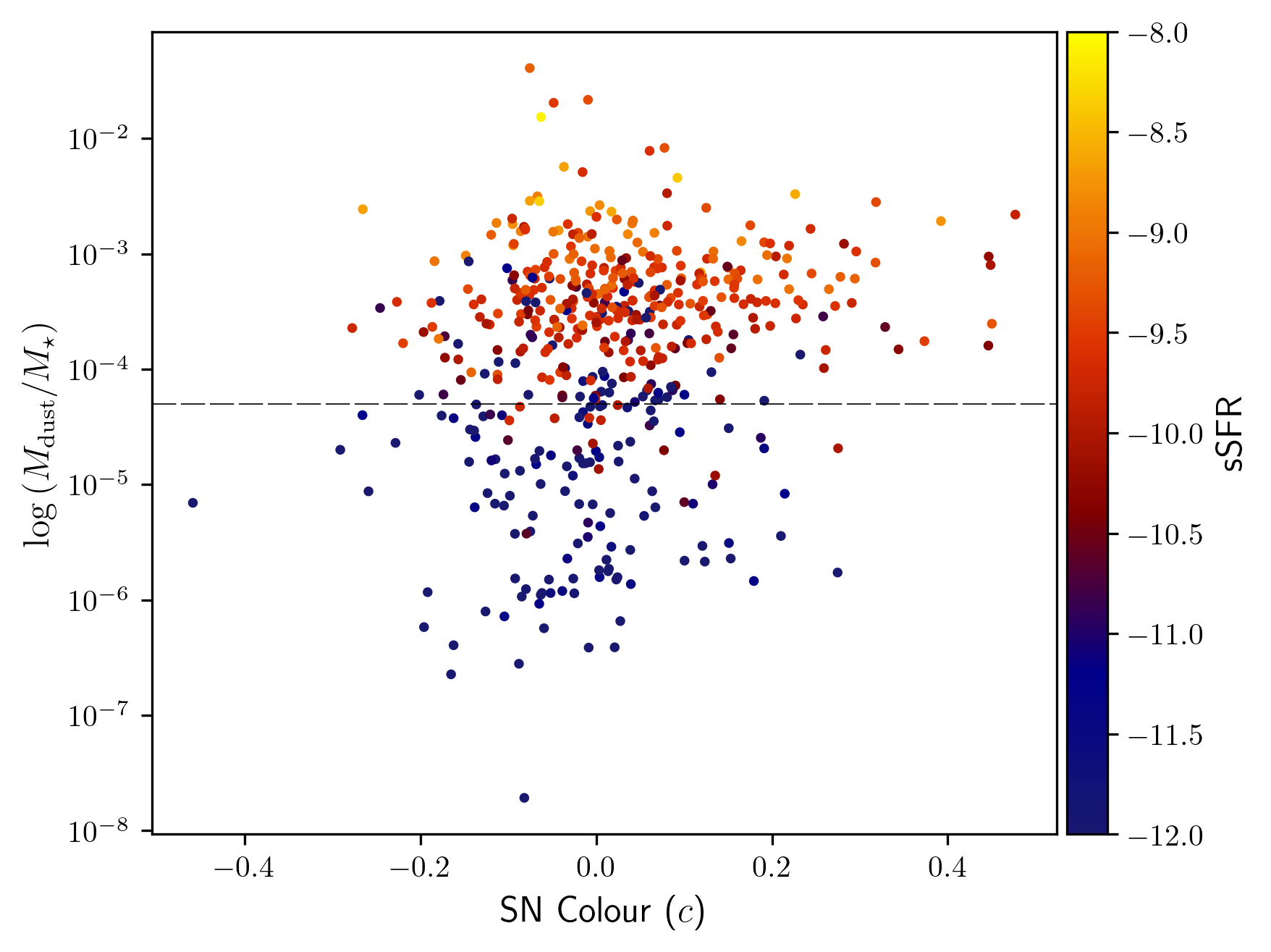} 
  \end{subfigure}
  \caption{SN colour, $c$, as a function of global host galaxy dust attenuation, $A_V^{\mathrm{HS}}$, (left) and specific dust mass, $M_{\mathrm{dust}}/M_\star$, (right). The samples are both colour-coded as a function of specific star-formation rate (sSFR). The horizontal dashed lines indicate the value of $A_V$ and/or $M_\mathrm{dust}$ that minimise the KS-test $p-$value.} 
  \label{SNc} 
\end{figure*}

The distributions of SN colour as a function of both the aforementioned dust parameters are displayed in Fig. \ref{SNc}. We note that the reddest SNe Ia are hosted by galaxies that are characterised by higher levels of dust attenuation and specific dust masses. Perhaps unsurprisingly as, for a globally dusty galaxy, the probability of a SN lying along a line of sight that coincides with a pocket of dust increases and the presence of dust preferentially acts to redden the light from a source.  

Mapping both samples as a function of specific star-formation rate (sSFR) shows that there exists a positive correlation with both dust parameters i.e., higher sSFR galaxies are also the dustiest. A strong positive correlation between SFR, $A_V$ and mass (or some pairwise combination of the three) has also been found in various literature for samples of star-forming galaxies (e.g., \citealp{1998ARA&A..36..189K,1998ApJ...498..579G,2000ApJ...533..682C, 2001MNRAS.327..697D,2010MNRAS.409..421G, 2013ApJ...763...92Z}). In the context of sSFR, SNe Ia appear to separate into two distinct populations in Fig.~\ref{SNc}: (i) those in low sSFR galaxies, which are generally bluer (i.e., with fewer red SNe) and span the full specific dust mass range; and (ii) those in high sSFR galaxies, which are invariably associated with high specific dust masses, with SNe that span the full -- and even broader -- colour range.

\begin{figure*} 
  \begin{subfigure}[b]{0.49\linewidth}
    \centering
    \includegraphics[width=1\linewidth]{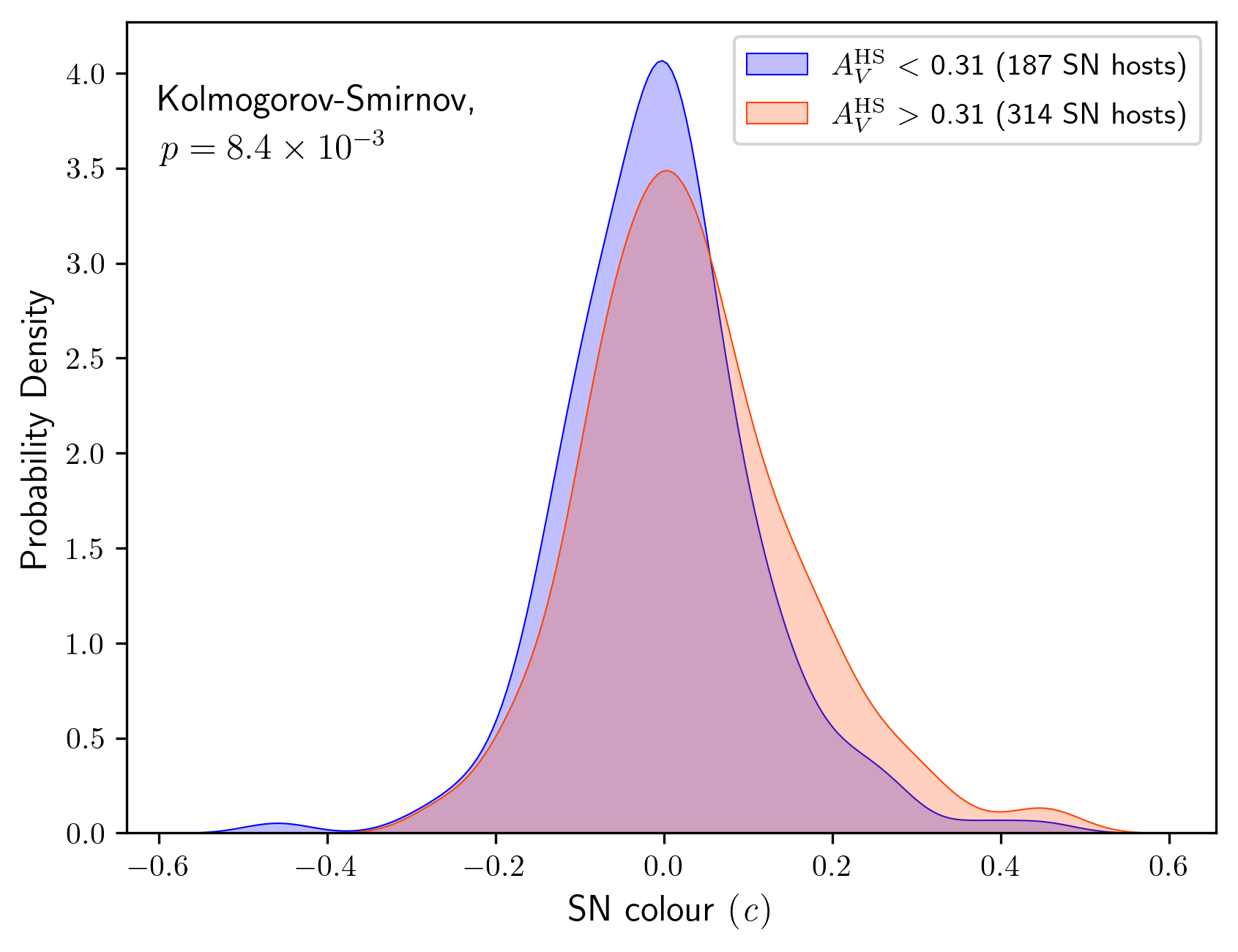} 
  \end{subfigure} \hfill
  \begin{subfigure}[b]{0.49\linewidth}
    \centering
    \includegraphics[width=1\linewidth]{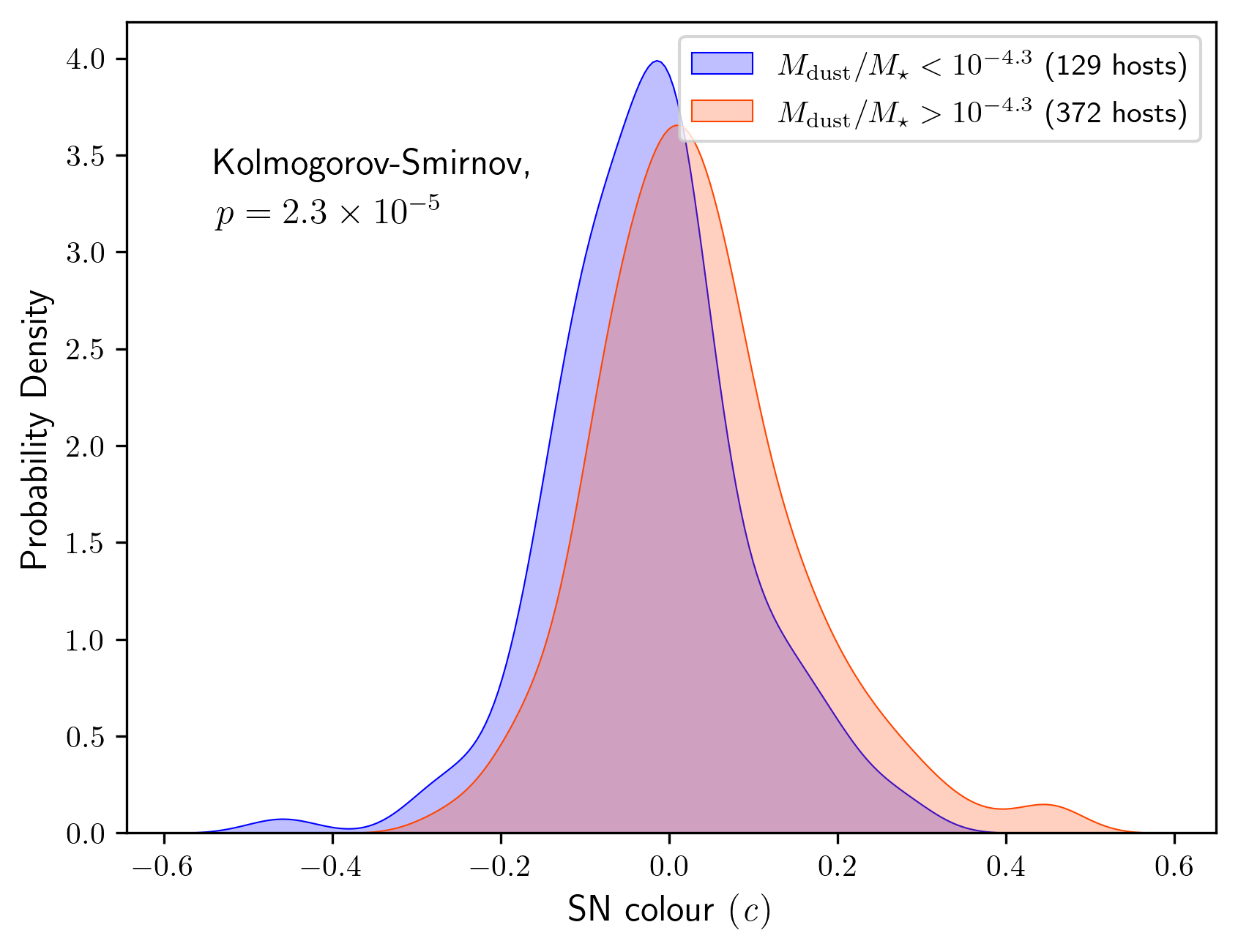} 
  \end{subfigure}
  \caption{Left: SN SALT3 colour, $c$  distributions for SNe in `low-attenuation' ($A_V^{\mathrm{HS}} \lesssim 0.31$) and `high-attenuation' ($A_V^{\mathrm{HS}} > 0.31$) host galaxies. Right: As left but for specific dust mass split at $M_{\mathrm{dust}}/M_\star = 10^{-4.3}$.} 
  \label{pdfs} 
\end{figure*}

To gauge whether it is dust attenuation or specific dust mass that is more strongly correlated with SN colour, we perform a two-sample statistical test to assess whether SN colour distributions differ for `low-' vs `high-attenuation/specific dust mass' galaxies. We split our `gold' sample of (501) galaxies at $A_V^{\mathrm{HS}} = 0.31$ mag and, separately, at $M_{\mathrm{dust}}/M_\star = 10^{-4.3}$. These values are chosen from minimising the $p$-value returned from a two-sample Kolmogorov-Smirnov (KS) test. The SALT3 $c$ distributions are displayed in Fig. \ref{pdfs}. A KS test indicates that it is highly statistically significant that the SALT3 $c$ distributions for SNe split by both dust parameters are sampled from different parent populations, with $p = 8.4 \times 10^{-3}$ and $p = 2.3 \times 10^{-5}$ for low- and high-attenuation and specific dust mass hosts, respectively.  The split on specific dust mass is the most significant, suggesting that this has a larger contribution than global host galaxy dust attenuation on SN colour variations. Again, unsurprising as a higher specific dust mass indicates a larger fraction of the total galaxy mass is comprised of dust, thus increasing the chance of SN light passing through a dusty sightline. However, we emphasise that the difference in SN colour is small.

It is worth commenting that the KS test difference for a split on $A_V$ is due to a broader red tail for SN Ia colours in high-$A_V$ hosts. Whereas for the split on specific dust mass, the difference is due to a shift in the entire distribution of SN Ia colours to redder values for SNe in high specific dust mass hosts.

\section{Results III: Splitting the SN Ia sample by host galaxy property -- Star-forming Main Sequence} \label{results3}

At the core of galaxy evolution studies is the effort to understand the formation and assembly of stellar mass ($M_\star$) content across cosmic time. In mind of this goal, coupled with the acquisition of statistically significant samples of data, one avenue of research has focused on exploring how galaxies occupy the SFR -- $M_\star$ plane. This representation into the evolutionary state of a galaxy reveals that, in the most basic sense, galaxies can be categorised into two distinct populations: actively star-forming versus quiescent, passive systems. Of the fraction that are star-forming (SF), a tight correlation is observed in the context of this plane -- a relation now commonly designated the main sequence (MS) of star-forming galaxies (SF-MS; \citealp{2004MNRAS.351.1151B, 2007ApJ...660L..43N, 2007A&A...468...33E, 2007ApJ...670..156D}). The SF-MS exists across a range of redshifts ($0 < z \lesssim 4$) and is usually characterised as having a slope of approximately unity, with little dispersion ($\sim 0.2 - 0.3$ dex) around the MS (though refer to the literature for a more extensive discussion). Under a mathematical formalism, the SF-MS can be modelled as a second-order polynomial of the form (as defined in \citealp{2012ApJ...754L..29W}), 

\begin{equation} \label{whit}
    \mathrm{log_{10}}[\mathrm{SFR}(z)] = \alpha(z)[\mathrm{log_{10}}(M_\star) - 10.5] + \beta(z)
\end{equation}

where $\alpha(z)$ and $\beta(z)$\footnote{These $\alpha$ and $\beta$ parameters are not to be confused with those used to characterize the SN stretch- and colour-luminosity relation.} parametrise the slope and normalisation of the SF-MS respectively and are given by,

\begin{equation*} 
\begin{array}{c}
    \alpha(z) = \alpha_1 + \alpha_2z \\[1em]
    \beta(z) = \beta_1 + \beta_2z + \beta_3z^2
\end{array}
\end{equation*}

with fit parameters: $\alpha_1$, $\alpha_2$, $\beta_1$, $\beta_2$ and $\beta_3$. We take the median values presented in Table~3 in \cite{2015MNRAS.453.2540J}, who also use \textit{Herschel} and \textit{Spitzer} data, in addition to optical and NIR wavelength coverage comparable to that used in this work.

\begin{figure*}
    \centering
    \includegraphics[width = 178mm]{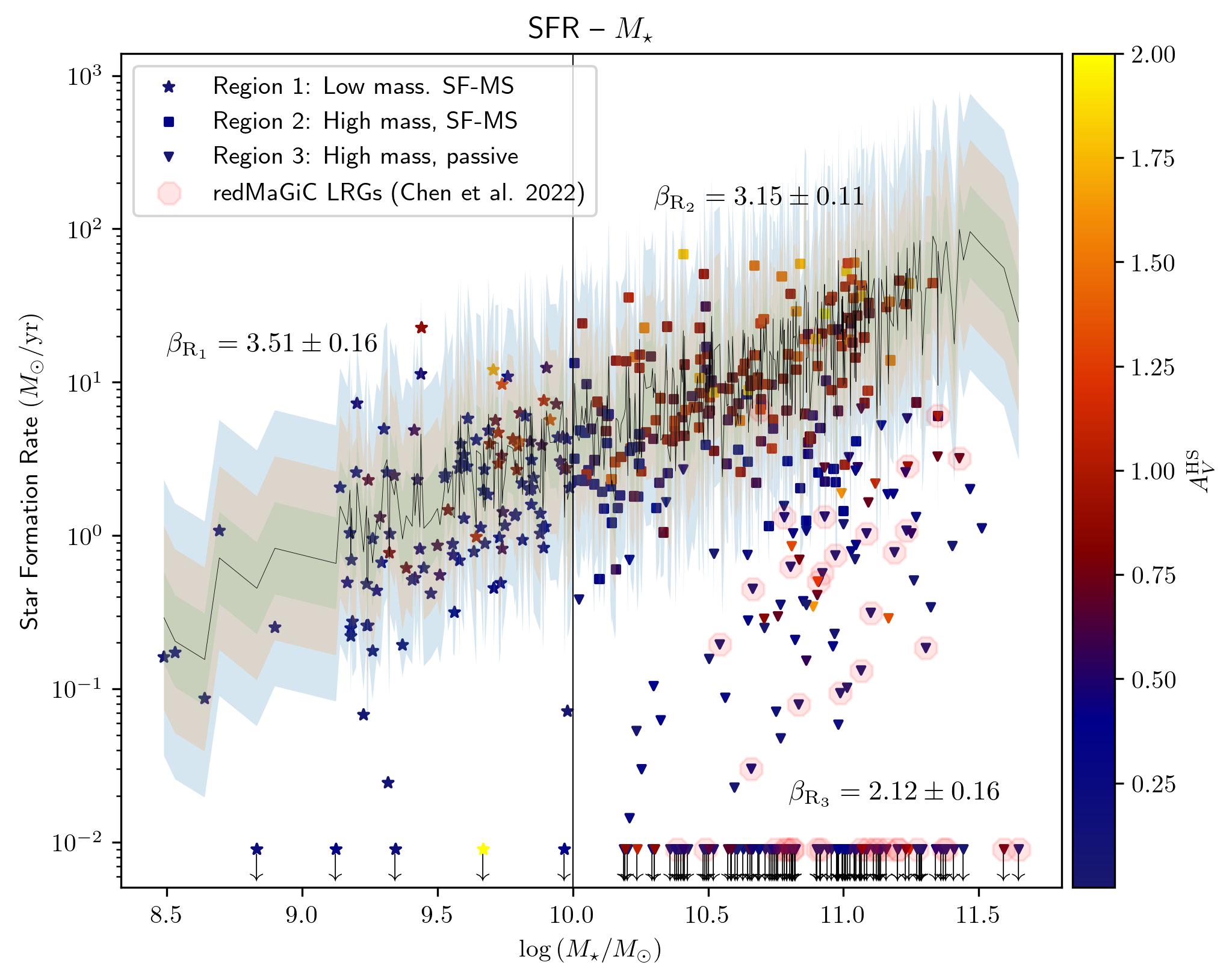}
    \caption{The distribution of SN Ia host galaxies across the SFR -- $M_\star$ plane. The plane is divided into three regions (coded by marker type) to isolate galaxies that are characterised by similar properties. Low mass galaxies are indicated by stars, where we apply an arbitrary cut on mass at the location of the "mass step" i.e., $ < $ 10$^{10}M_\odot$. High mass galaxies ($ > $ 10$^{10}M_\odot$) are then split into two populations according based on whether they lie within a 3$\sigma$ interval of the SF-MS (Eq. \ref{whit}), which is highlighted by the blue shaded region. To aid the interpretation of our nuisance parameter constraints for different $\sigma$ cuts, which are presented in Table \ref{beta-vals}, we also plot the 1$\sigma$ (green shaded region) and 2$\sigma$ (orange shaded region) intervals. High mass, high SFR galaxies are shown as squares and conversely, high mass, low SFR (passive) galaxies are shown as triangles. Galaxies with SFR values below 0.01 $M_\odot/\mathrm{yr}$ are set as upper limits to this value and are indicated by downward arrows for illustration purposes. The points are colour-mapped by host galaxy dust attenuation, $A_V^{\mathrm{HS}}$, constrained when including \textit{Herschel} and \textit{Spitzer} data. The dynamic range on the colour bar is altered to limit the upper bound on $A_V^{\mathrm{HS}}$ to 2 mag, to emphasise differences between host attenuation across the SFR -- $M_\star$ plane. Luminous Red Galaxies (LRGs) in the redMaGiC sample from \citealp{Chen_2022} (Section \ref{redmag_chen}) that overlap with our `gold' sample are displayed by red hexagonal contours. Refer to Table. \ref{beta-vals} for the full summary of cosmological SN nuisance parameter constraints.}
    \label{sfms}
\end{figure*}

In the interest of studying correlations between SN Ia properties and their host galaxy environments, we use the parametric form of the SF-MS (Eq. \ref{whit}) to split the SFR -- $M_\star$ plane into three regions, each of which is comprised of a population of hosts that share common characteristics. The first region consists of SNe Ia located in low mass host galaxies, which we arbitrarily define as a stellar mass cut that is coincident with the position of the "mass step" i.e., $ < $ 10$^{10}M_\odot$. We do not explore other mass cuts in this analysis and note that the significance of our results is expected to differ based on the split point used (refer to \citealp{2010MNRAS.406..782S}, Table 5). Next, we use SFR as a proxy to classify the high mass hosts ($ > $ 10$^{10}M_\odot$) into two populations. We split host galaxies into those on and off the main sequence by cutting those more than three times the scatter from the SF-MS. In brief, the second region is defined by a cut $\Delta$log$_{10}$(SFR) < 3$\sigma$, isolating host galaxies that lie on the SF-MS (note this applies only to galaxies 3$\sigma$ below the SF-MS, not above it). We take $\sigma = 0.3$ dex following \citealp{2015MNRAS.453.2540J} (hence 3$\sigma = 0.9$ dex). The third and final region, with $\Delta$log$_{10}$(SFR) > 3$\sigma$, encompasses SN host galaxies that have evolved off and lie below the SF-MS and are thus, characterised by low SFRs i.e., a passive, quenched population. The choice to separate high-mass star-forming and high-mass passive hosts by whether they lie within a 3$\sigma$ confidence interval around the SF-MS is quite conservative, ensuring that we minimise the number of star-forming galaxies in the region that we aim to limit to passive galaxies. We explore adopting other $\sigma-$cuts when splitting our sample in this context but retain a 3$\sigma$ cut as the basis for our main analysis. This choice is made prior to examining the effects of alternative $\sigma-$cuts, which we include in this work solely for comparison, and is not guided by any attempt to present the most optimal results. The distribution of SN host galaxies in the triply divided SFR -- $M_\star$ plane is presented in Fig. \ref{sfms}. In the following, we explore various properties and nuisance parameter constraints for SNe Ia in each of the three aforementioned regions.

\subsection{Nuisance parameter constraints}

\begin{figure*}
    \centering
    \includegraphics[width = 180mm]{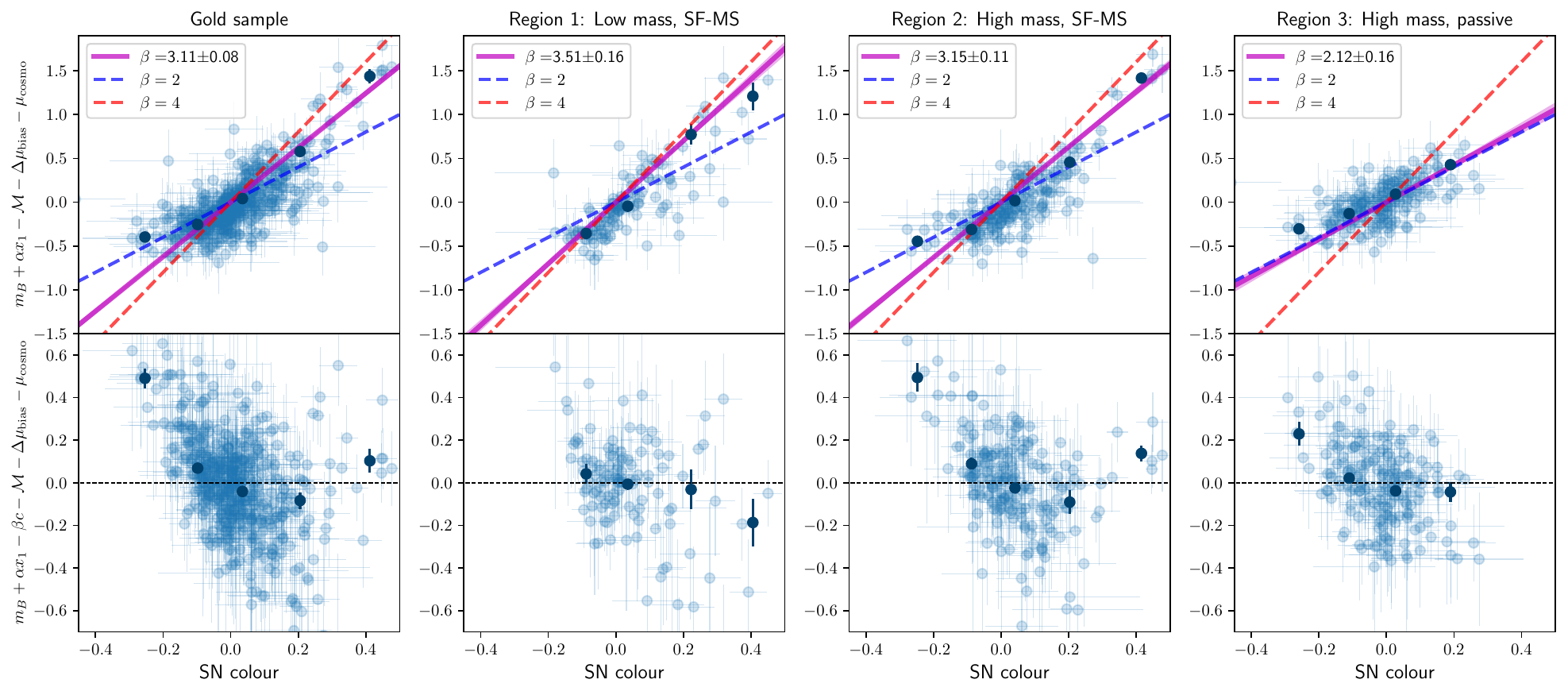}
    \caption{SN SALT3 colour vs. SN standardized brightness (see eq.~\ref{eq:SNstand}) calculated both excluding (top panel) and including (bottom panel) colour-dependent corrections ($-\beta c$ term), and subtracting \textit{Planck}-like cosmology ($\mu_\mathrm{cosmo}$) i.e., Hubble residuals. We present the best-fit $\beta$ values for four different samples (from left to right): the gold sample and the Regions 1, 2 and 3 sub-samples (see Section \ref{results3}). $\beta$ slopes of 2 and 4, which are the extremes of values found in the literature, are overplotted to guide the eye.}
    \label{betaplot2}
\end{figure*}

\begin{figure}
    \centering
    \includegraphics[width = 85mm]{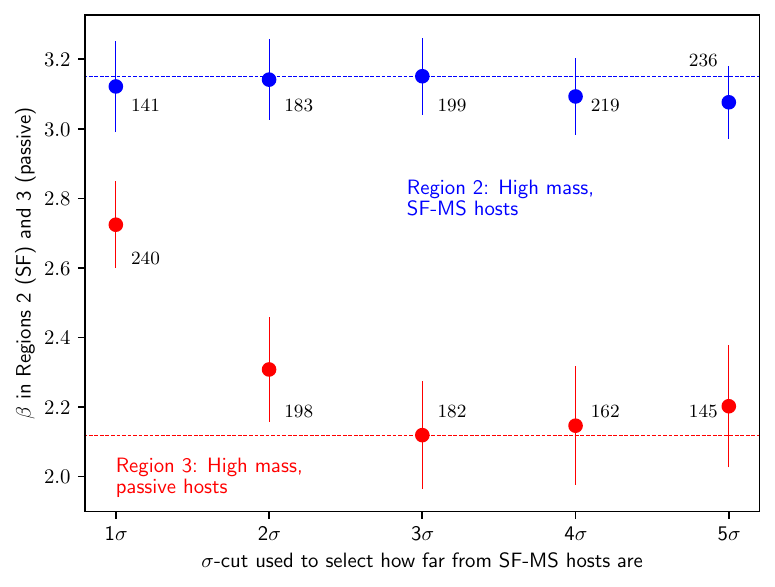}
    \caption{Best-fit $\beta$ using SNe in region 2 (blue) and region 3 (red), shown as a function of $\sigma$-cut used to select SF-MS vs passive host galaxies. We test a range of $\sigma$-cuts, from the least conservative selection of passive hosts (hosts that are 1$\sigma$ away from SF-MS are considered to be passive, i.e., assigned to region 3) to the strictest selection of passive hosts (only hosts that are at least 5$\sigma$ away from SF-MS are considered to be passive, i.e., assigned to region 3). For reference, the nominal $\beta$ values (measured using a $3\sigma$ cut) are shown as horizontal dashed lines for the two regions. Next to each datapoint we display the number of SNe in each regions $\sigma$-cut sample.}
    \label{beta_vs_sigma}
\end{figure}

In this section we investigate the SN nuisance parameters ($\alpha$, $\beta$) that are used to correct for variations in SN luminosities. These parameters are measured using BBC (see Section~\ref{BBC}). We obtain these for both our total `gold' sample and for the subsamples of SNe Ia in each division of the SFR -- $M_\star$ plane to check the level of consistency when splitting on host galaxy properties. We also explore the absolute magnitude of SNe Ia ($M$ in Eq. \ref{eq:SNstand}). We note, however, that without an absolute calibrated distance scale, $M$ is degenerate with $H_0$. Instead, cosmological fits marginalise over a single parameter, $\mathcal{M} = M + \mathrm{log}_{10}(c/H_0) + 25$, that combines these terms. Therefore, we do not present our results in terms of $M$, which requires an assumption for the value of $H_0$. Instead, we report our findings in terms of $\Delta \mathcal{M}$, the relative value of $\mathcal{M}$ with respect to our `gold' sample constraint. This choice does not impact our results in any way. We present our results in Table \ref{beta-vals} and our findings are as follows,

\begin{enumerate}[label=(\roman*), leftmargin=*,align=left,labelindent=\parindent, labelsep=0.5em, itemindent=!]

\item The values of $\alpha$ in each of the three sub-regions are consistent, with the largest difference at the ${\sim}0.5\sigma$ level (between regions 1 and 3). 

\item SNe Ia in low mass hosts (region 1) are characterised by higher values of $\beta$ than those in high mass hosts. The largest difference (${\sim}6.1\sigma$) exists between regions 1 and 3 (passive hosts). Smaller differences in $\beta$ are seen between regions 1 and 2 (high mass, MS), significant at the ${\sim}1.9\sigma$ level. These differences persist even amongst high mass hosts (regions 2 and 3), with passive hosts displaying the smallest $\beta$ values (${\sim}5.3\sigma$ difference). We plot the Hubble residuals, excluding the colour-dependent correction ($-\beta c$) term, vs. SALT3 SN colour in Fig. \ref{betaplot2} (top panel), and overlay the best fit lines whose slopes are given by our measured values of $\beta$ for each region. For completeness, Hubble residuals with the colour correction included are plotted in the bottom panel of Fig. \ref{betaplot2}. The observed trend of positive residuals for blue SN colours is expected and arises due to parameter migration effects (see \citealp{Scolnic_2016} for a detailed explanation).

\item When restricting our sample to $z\sim0.6$ to mitigate the effects of selection biases, the difference in $\beta$ values between the three regions is still significant and follow the same trend observed for the full sample (region 1 has the highest $\beta$, while region 3 the lowest).  We note that this redshift cut results in a $\sim1\sigma$ increase in $\beta$, on average, across all best fits. This is not surprising as selection effects primarily affect highly reddened (hence faintest) SNe, thus resulting in a flattening of the colour-luminosity relation parametrized by $\beta$.

\item We test different sigma cuts (Fig. \ref{beta_vs_sigma}) from the SF-MS when defining regions 2 and 3 and measure how this impacts the value of $\beta$ in the two regions. When applying a higher sigma cut, region 3 is representative of a `purer' sample of passive hosts and the measured value of $\beta$ is consistently lower (with the largest difference between regions up to a ${\sim}6\sigma$ significance level). Only for a $1\sigma$ cut does the significance of the $\beta$ difference between region 2 and region 3 decrease below $3\sigma$. This is likely due to region 3 being contaminated by SF-MS galaxies.
The $\beta$ value for region 2 is consistent for different sigma cuts.

\item Consistent with previous works \citep{2010ApJ...715..743K,2010MNRAS.406..782S,2010ApJ...722..566L}, we find SNe Ia in high mass hosts (regions 2 and 3) are, on average, brighter (negative $\Delta\mathcal{M}$) post-standardization than those in low mass hosts. This well-known trend is referred to as the \lq mass step\rq\ and has been confirmed in previous cosmological analyses that fit the full SN Ia sample. The mass step is usually defined as the difference between the average Hubble residuals for low- and high-mass galaxies. As we are minimising the residuals for each region separately, we instead look at differences in the SN absolute rest-frame magnitude  across the three regions. The most significant brightness differences ($\sim 0.121$ mag) are seen between SNe in low mass (region 1) and high mass, passive (region 3) hosts, at a ${\sim}5.8\sigma$ significance level. The smallest differences in brightness ($\sim 0.055$ mag) are between SNe in low mass and high mass, MS (region 2) hosts (${\sim}2.6\sigma$). Amongst both high mass regions (2 and 3), differences persist at a ${\sim}3.7\sigma$ significance level. We highlight that these differences in $\Delta\mathcal{M}$ and/or $M$ come after having applied a separate, optimised $\beta$ correction to each subsample.

\item SNe Ia in high mass hosts display smaller r.m.s scatter in the Hubble residuals than low mass hosts. The r.m.s scatter amongst SNe in both high mass regions are similar. For the $z$-cut sample, which is a more `complete' sample that is less affected by selection biases, the r.m.s decreases when moving in order from regions 1--3 (which is the same trend seen as for the nominal sample except the differences are more significant). Region 3, which has the lowest r.m.s of $0.151 \pm0.025$, is also characterised by the lowest value of $\beta$. This may be suggestive of SNe in passive hosts forming the most homogeneous sample due to having a combination of both the lowest $\beta$ and lowest r.m.s.

\end{enumerate}

\begin{table*} 
	\centering
	\caption{Nuisance parameter values constrained using BBC for SNe Ia in host galaxies that occupy three distinct regions of the SFR-$M_\star$ plane. `Nominal' indicates we use all SNe in a given subregion without applying any redshift cuts, while \lq $z<0.6$\rq\ indicates the redshift cut we apply to minimise the effect of survey selection biases on our sample. (see Sec.~\ref{BBC}). Moreover, \lq $2\sigma$-cut\rq\ and \lq $4\sigma$-cut\rq\ indicate the different cuts applied to select passive vs. SF-MS hosts. In reference to the nominal results for regions 2 and 3, we use a $3\sigma$ cut.}
	\label{beta-vals}
	\begin{tabular}{l l c c c c c}  
		\hline
		\hline
		\textbf{SF-MS Region} & \textbf{Selection Cuts} & $N_\mathrm{SN}$$^{(*)}$ & \textbf{$\alpha$} &\textbf{$\beta$} & $\Delta \mathcal{M}$$^{(\dag)}$ & r.m.s $^{(\ddag)}$\\
        \hline
\textbf{Gold sample} & Nominal & 495 & 0.157$\pm$0.009 & 3.11$\pm$0.08 & -- & 0.257$\pm$0.016 \\
 & ($z<0.6$) & 238 & 0.166$\pm$0.011 & 3.30$\pm$0.09 & 0.005$\pm$0.012 & 0.182$\pm$0.017 \\
\hline
 \textbf{Region 1:} Low mass, SF-MS ($<10^{10}M_\odot$) & Nominal & 116 & 0.185$\pm$0.023 & 3.51$\pm$0.16 & 0.055$\pm$0.019 & 0.277$\pm$0.037 \\
 & ($z<0.6$) & 70 & 0.182$\pm$0.023 & 3.64$\pm$0.15 & 0.062$\pm$0.019 & 0.180$\pm$0.031 \\
\hline
 \textbf{Region 2:} High mass ($>10^{10}M_\odot$), SF-MS ($\Delta$log$_{10}$(SFR)<3$\sigma$) & Nominal & 197 & 0.180$\pm$0.019 & 3.15$\pm$0.11 & 0.000$\pm$0.015 & 0.238$\pm$0.024 \\
 & ($z<0.6$) & 92 & 0.208$\pm$0.024 & 3.30$\pm$0.13 & -0.017$\pm$0.018 & 0.165$\pm$0.024 \\
& (2$\sigma$-cut) & 181 & 0.178$\pm$0.019 & 3.14$\pm$0.12 & 0.000$\pm$0.015 & 0.241$\pm$0.025 \\
& (4$\sigma$-cut) & 217 & 0.184$\pm$0.018 & 3.09$\pm$0.11 & -0.005$\pm$0.014 & 0.242$\pm$0.023 \\
\hline
 \textbf{Region 3:} High mass ($>10^{10}M_\odot$), passive ($\Delta$log$_{10}$(SFR)>3$\sigma$) & Nominal  & 182 & 0.172$\pm$0.013 & 2.12$\pm$0.16 & -0.066$\pm$0.014 & 0.207$\pm$0.022 \\
 & ($z<0.6$) & 76 & 0.181$\pm$0.017 & 2.36$\pm$0.21 & -0.052$\pm$0.018 & 0.151$\pm$0.025 \\
& (2$\sigma$-cut) & 198 & 0.171$\pm$0.014 & 2.31$\pm$0.15 & -0.055$\pm$0.014 & 0.215$\pm$0.022 \\
& (4$\sigma$-cut) & 162 & 0.170$\pm$0.013 & 2.15$\pm$0.17 & -0.066$\pm$0.015 & 0.211$\pm$0.024 \\
		\hline
	\end{tabular}
         {\raggedright $^{(*)}$ The number of SNe Ia included in the BBC fit include a $4\sigma$ outlier rejection. For this reason, the BBC fit of the `gold' sample (501 SNe, see Table~\ref{numbers}), for example, only includes 495 SNe. \\
         $^{(\dag)}$ $\Delta\mathcal{M}$ is the difference in absolute brightness compared to the full `gold' sample defined in this work, with the errors given in quadrature. The quoted uncertainties on $\Delta\mathcal{M}$ should be regarded as a conservative upper bound as the assumption of statistical independence -- which underlies the quadrature addition of uncertainties -- is not strictly valid in this case (each region sample is a subset of the gold sample which means the error on $\mathcal{M}$ for the gold sample and any region subsample are correlated). The true uncertainty on $\Delta\mathcal{M}$ is expected to be lower.  \\
         $^{(\ddag)}$ Sample (or sub-sample) Hubble residuals r.m.s. scatter.\\
         \par}
    \end{table*}

\subsection{Distribution of SN Ia properties}

\subsubsection{SN colour}

Here, we explore the distribution of SN Ia properties in each region of the SFR--$M_\star$ plane. Fig.~\ref{cx1} (left) shows the colour distributions of SNe Ia in each of the three regions. We do not find any strong trends between SN colours in low- and high-mass stellar hosts, nor have any been found to date in the literature. The mean of SN colours in region 2 (high mass, MS) display a shift toward redder colours. This aligns with the $A_V^{\mathrm{HS}}$ estimates associated with this region. Fig. \ref{sfms} (or, alternatively, mapping the plane by specific dust mass in Fig.~\ref{Mdsfms}) shows that these host galaxies are more affected by dust attenuation (as expected for highly star-forming systems), which may redden the SNe (see Section~\ref{results2}, Fig. \ref{pdfs}). 

The most statistically significant difference between SN colour distributions is between regions 2 and 3, with a KS test returning $p = 3.4 \times 10^{-4}$. Perhaps unsurprisingly as galaxies in these regions display the biggest differences in dust attenuation and specific dust masses. \cite{2025A&A...694A...4G} also find higher mass galaxies to host redder SNe, with a return to bluer SN colours in the most massive (i.e., our region 3) noted by both \cite{2023MNRAS.519.3046K} and \cite{2025A&A...694A...5P}. We note that in our analysis, we constrain the global host galaxy $A_V$ and, as a result, the role that circumstellar dust and/or line of sight attenuation has on producing the observed SN colour distribution remains unclear.
The colour distributions in Fig. \ref{cx1} are produced using a $3\sigma-$cut to separate regions 2 and 3 (high-mass, SF-MS vs. high-mass, passive). As an additional test, we reproduce our results using other $\sigma-$cuts, the percentiles for which are given in Table~\ref{tab:colour-percentiles}. There are no significant changes to the resultant colour distributions.

\begin{table}
    \centering
    \begin{tabular}{|l|l|c|c|c|}
        \hline
        \hline
        \textbf{SF-MS Region} & \textbf{Selection Cuts} & 16$^{\mathrm{th}}$ & 50$^{\mathrm{th}}$ & 84$^{\mathrm{th}}$ \\
        \hline
        \multirow{1}{*}{Region 1} & Nominal & -0.07 & 0.01 & 0.16 \\
        \hline
        \multirow{3}{*}{Region 2} & Nominal & -0.08 & 0.03 & 0.16 \\
        & 2$\sigma$-cut & -0.06 & 0.03 & 0.16 \\
        & 4$\sigma$-cut & -0.08 & 0.03 & 0.15 \\
        \hline
        \multirow{3}{*}{Region 3} & Nominal & -0.12 & -0.01 & 0.07 \\
        & 2$\sigma$-cut & -0.12 & -0.01 & 0.08 \\
        & 4$\sigma$-cut & -0.12 & -0.02 & 0.07 \\
        \hline
    \end{tabular}
    \caption{Percentiles of SN \textbf{colour} as a function of SF-MS region and selection cuts. `Nominal' refers to the baseline 3$\sigma$ cut. Region 1 is only a cut on mass and is unaffected by $\sigma-$cuts. We include the region 1 percentiles for completeness.}
    \label{tab:colour-percentiles}
\end{table}

\subsubsection{SN stretch}

\begin{figure*} 
  \begin{subfigure}[b]{0.49\linewidth}
    \centering
    \includegraphics[width=1\linewidth]{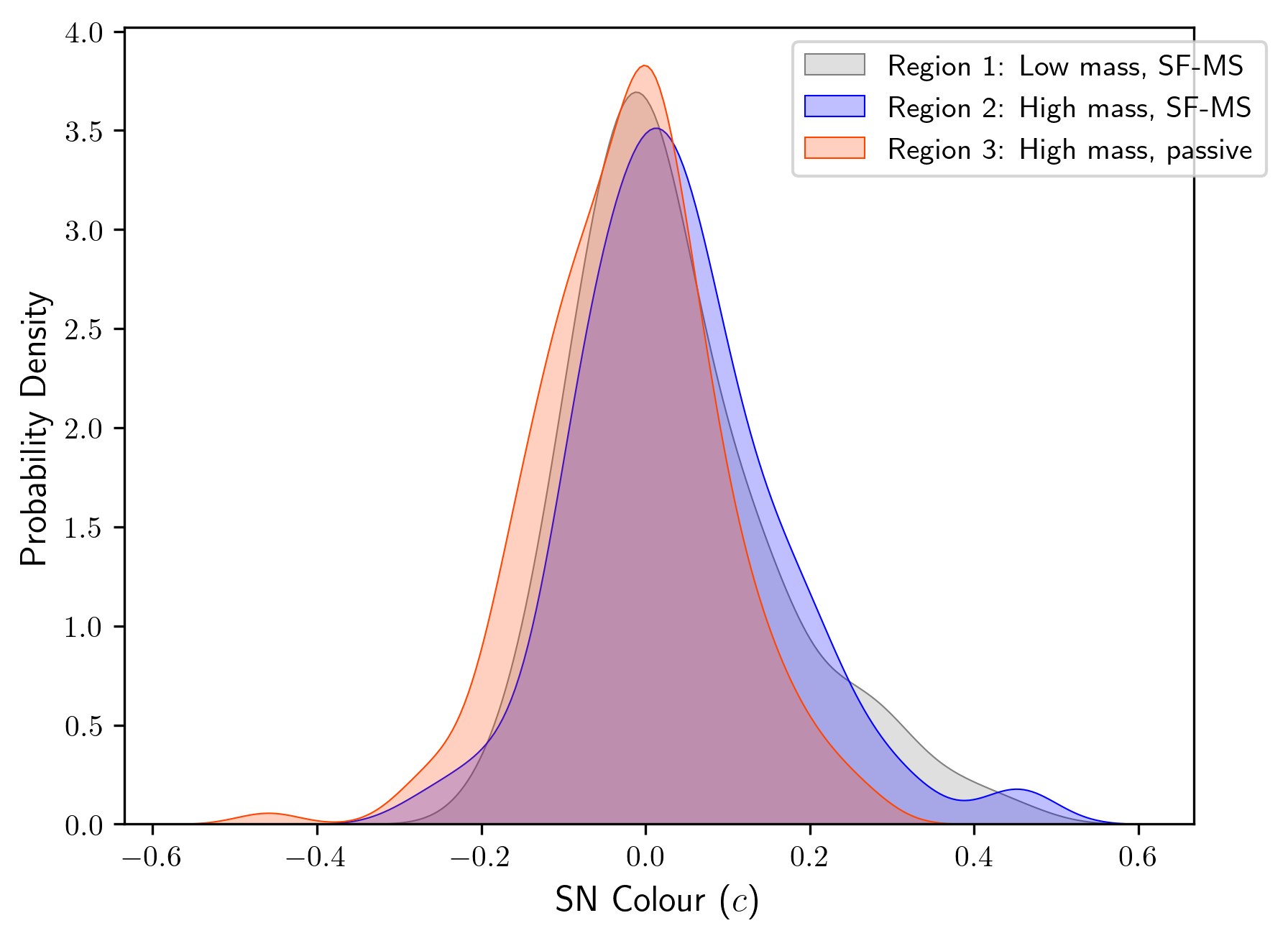} 
  \end{subfigure} \hfill
  \begin{subfigure}[b]{0.49\linewidth}
    \centering
    \includegraphics[width=1\linewidth]{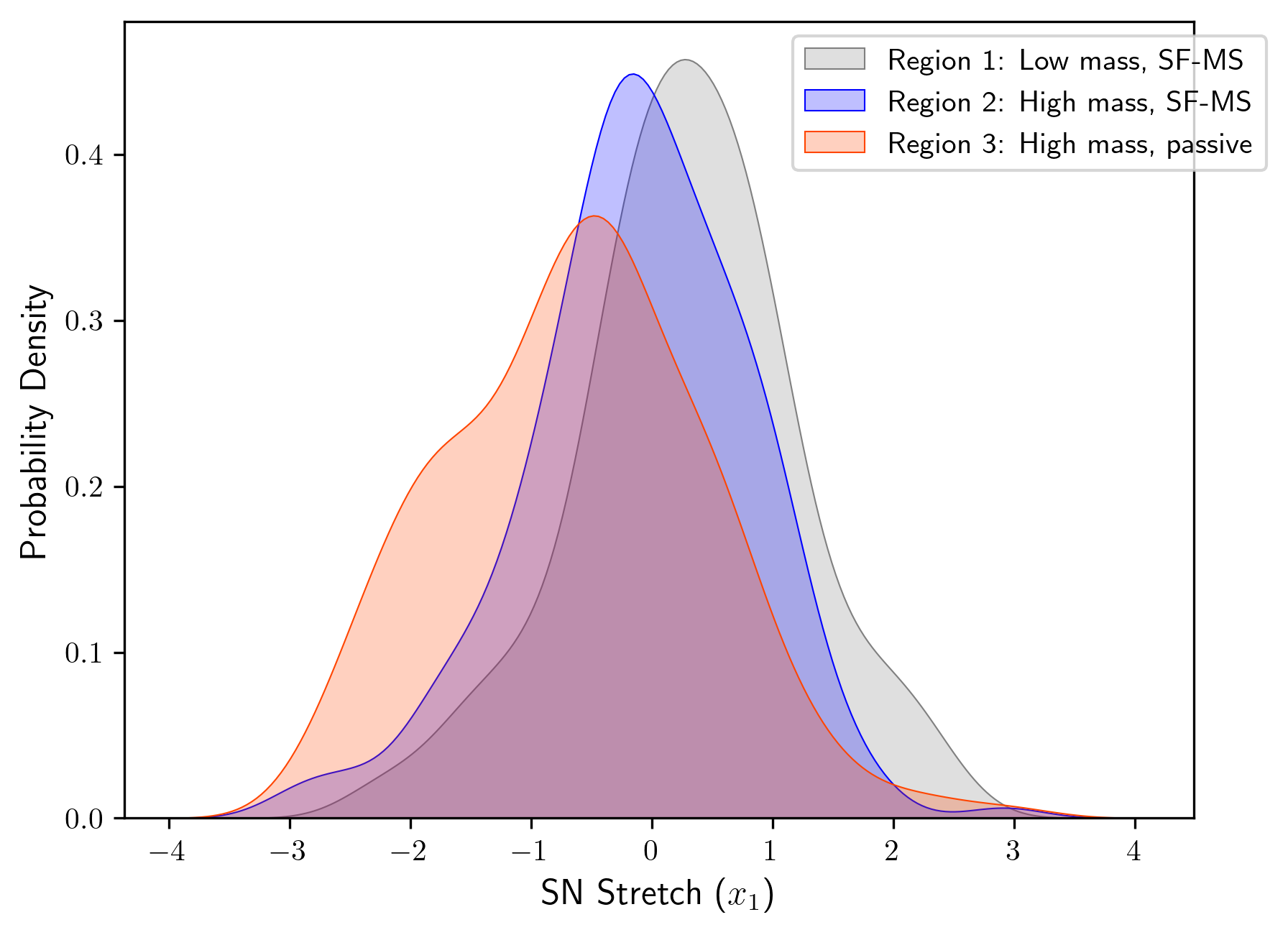} 
  \end{subfigure}
  \caption{Distribution of SN colour, $c$ (left) and SN stretch, $x_1$ (right) in each of the three regions of the SFR-$M_\star$ plane.} 
  \label{cx1} 
\end{figure*}

Correlations between SN stretch and host galaxy properties are amongst the more well-documented in the literature (\citealp{1996AJ....112.2391H, 1999AJ....117..707R,2000AJ....120.1479H,2006ApJ...648..868S}). In Fig. \ref{cx1} (right), we show the distributions of SN stretch in the regions 1, 2 and 3 and confirm the observed trends that early-type passive galaxies (i.e., region 3 in our analysis) primarily host low-stretch SNe. In a similar vein, the mean stretch of SNe in galaxies with active star-formation is shifted toward higher values, with a relative lack of low-stretch events (regions 1 and 2). Paraphrasing in terms of host galaxy stellar mass, low mass galaxies (region 1) tend to have an absence of low-stretch events (e.g., \citealp{2010MNRAS.406..782S}). As our host galaxy measurements allow us to split high mass galaxies into two regions (2 and 3), it appears that the SNe in passive hosts drive the stretch distribution to lower values of $x_1$. The shift of stretch distributions between regions aligns well with the model of \cite{2021A&A...649A..74N,2025A&A...695A.140G}. In this context, region 1 comprises only SNe from the `high-stretch' mode; region 2, with a combination of young and old stellar populations, is a mixture of low- and high-stretch mode SNe and region 3, with a \textit{pure} old stellar population, has a higher fraction of `low-stretch' mode SNe. The apparent negative shift of the entire distribution from region 1 to region 3 is also predicted in the \cite{2021A&A...649A..74N,2025A&A...695A.140G} model. As with the SN colour distributions, we test the effect of other $\sigma-$cuts on our results. The percentiles are presented in Table~\ref{tab:x1-percentiles}, where we observe no significant differences in the SN stretch distributions.

\begin{table}
    \centering
    \begin{tabular}{|l|l|c|c|c|}
        \hline
        \hline
        \textbf{SF-MS Region} & \textbf{Selection Cuts} & 16$^{\mathrm{th}}$ & 50$^{\mathrm{th}}$ & 84$^{\mathrm{th}}$ \\
        \hline
        \multirow{1}{*}{Region 1} & Nominal & -0.40 & 0.30 & 1.08 \\
        \hline
        \multirow{3}{*}{Region 2} & Nominal & -0.93 & -0.14 & 0.80 \\
        & 2$\sigma$-cut & -0.91 & -0.14 & 0.84 \\
        & 4$\sigma$-cut & -0.96 & -0.16 & 0.76 \\
        \hline
        \multirow{3}{*}{Region 3} & Nominal & -1.76 & -0.59 & 0.44 \\
        & 2$\sigma$-cut & -1.76 & -0.54 & 0.44 \\
        & 4$\sigma$-cut & -1.82 & -0.62 & 0.45 \\
        \hline
    \end{tabular}
    \caption{As Table \ref{tab:colour-percentiles} but for SN \textbf{stretch}.}
    \label{tab:x1-percentiles}
\end{table}

\section{Discussion} \label{discussion}

\subsection{The slope of the SN colour -- luminosity relation}
\label{sec:beta_constr}

\begin{figure*} 
  \begin{subfigure}[b]{0.49\linewidth}
    \centering
    \includegraphics[width=1\linewidth]{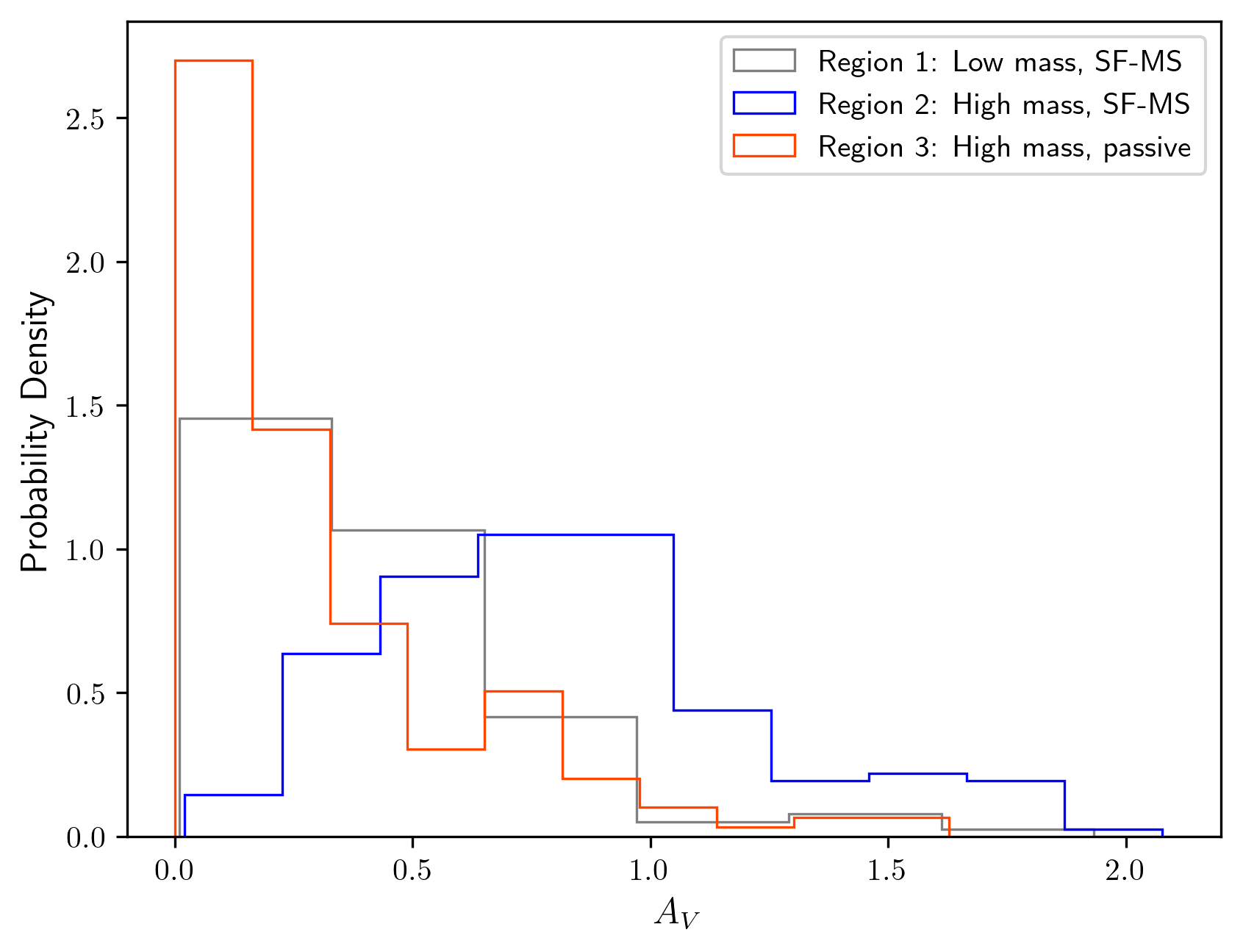} 
  \end{subfigure} \hfill
  \begin{subfigure}[b]{0.49\linewidth}
    \centering
    \includegraphics[width=1\linewidth]{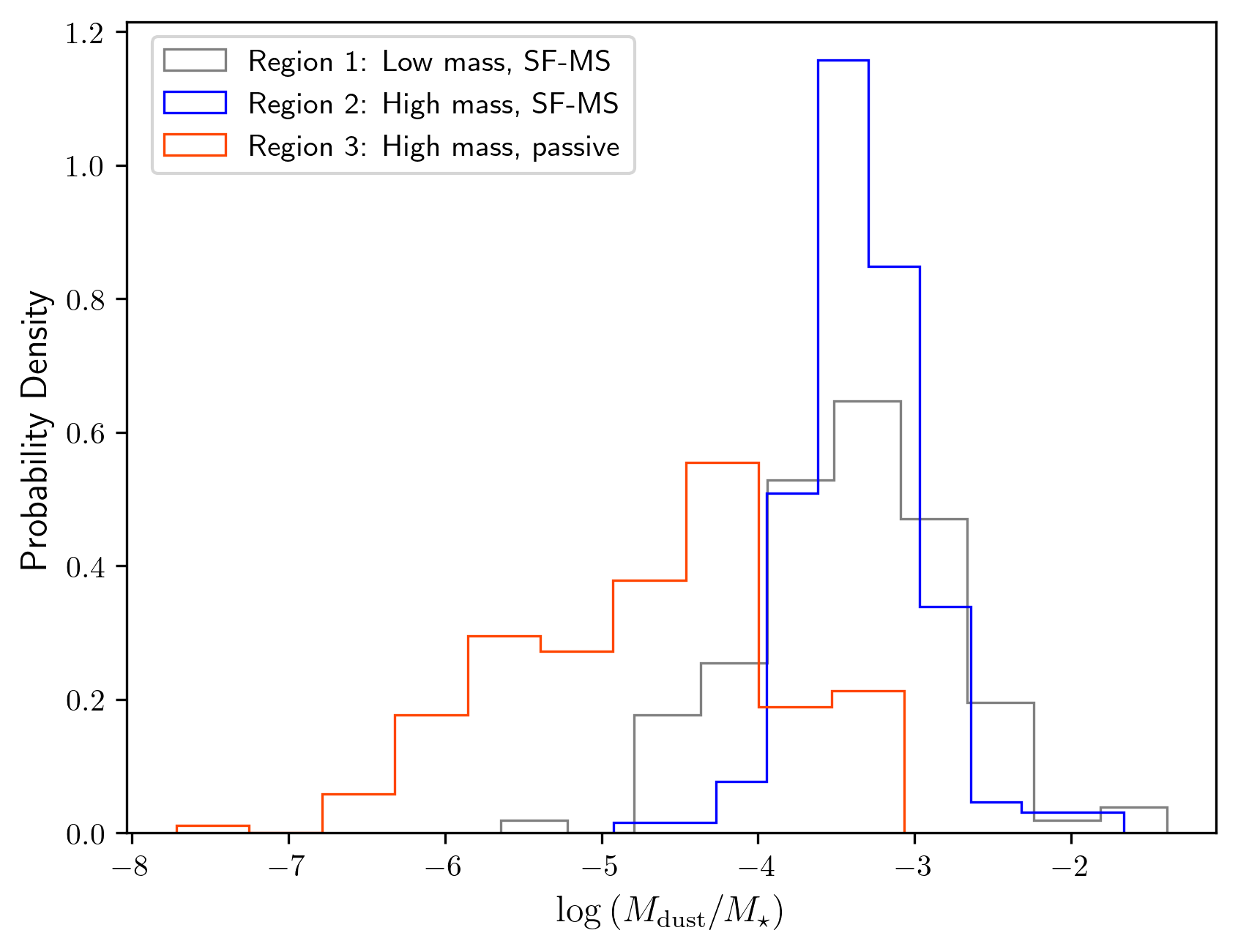} 
  \end{subfigure}
  \caption{Global dust attenuation, $A_V$ (left) and specific dust mass (right) distributions for host galaxies in each region of the triply divided SFR-$M_\star$ plane. We note that axis is truncated solely for visualisation purposes to exclude a single host galaxy in region 1 with $A_V > 3$ mag.} 
  \label{3regAvMd} 
\end{figure*}

Cosmological analyses that use the full SN Ia sample (i.e., not split by host properties) to measure the slope of the colour-luminosity relation generally recover values of $\beta$ ${\sim}$ 3. The recent DES-SN5YR results (\citealp{2024ApJ...975...86V}) find $\beta = 3.14 \pm 0.03$, which is consistent with our total `gold' sample results, where we measure $\beta = 3.11 \pm 0.08$.

As for the underlying scientific justification, the favoured hypothesis for what drives higher values of $\beta$ is a combination of intrinsic SN variations and dust. To gain further insight, we map each region on the SFR--$M_\star$ plane by both global host galaxy dust attenuation $A_V^{\mathrm{HS}}$ (Fig.~\ref{sfms}) and specific dust mass (Fig.~\ref{Mdsfms}). We remind the reader that host galaxy $R_V$ is not well constrained, even with our extensive multi-wavelength data (Section \ref{results}). The distributions of host galaxy dust attenuation and specific dust mass by region are presented in Fig.~\ref{3regAvMd}. We find that region 3 (high-mass, passive) hosts are characterised by the lowest values of dust attenuation and specific dust masses. SNe Ia in this region also yield the lowest value of $\beta = 2.12 \pm 0.16$. The measured $\beta$ for SNe Ia in star-forming host galaxies is higher, with $\beta = 3.51 \pm 0.16$ and $\beta = 3.15 \pm 0.11$ for regions 1 and 2, respectively. Host galaxies in these regions are also subject to higher levels of dust attenuation and have higher specific dust masses (Fig. \ref{3regAvMd}). 

One possible interpretation of these findings may be that $\beta$ for SNe Ia in region 3, where galaxies are the least globally dusty, is closer to the intrinsic SN $\beta$. In a similar vein, as host galaxies in regions 1 and 2 are globally dustier, the higher $\beta$ values for SNe Ia in these regions may be due to the effects of dust. Whilst a reasonable hypothesis, we refrain from making any conclusive remarks regarding the driver of SN $\beta$ variations in different host galaxy environments. This is because our dust parameter measurements are a global property of the galaxy, instead of locally to the position of the SN. Many works suggest that it is line-of-sight and/or circumstellar dust around the SN itself that drive variations in their luminosities (e.g., \citealp{2005ApJ...635L..33W,2010AJ....139..120F}). At present, our current analysis lacks the necessary constraints to validate or disprove this theory. 

\begin{figure}
    \centering
    \includegraphics[width = 78mm]{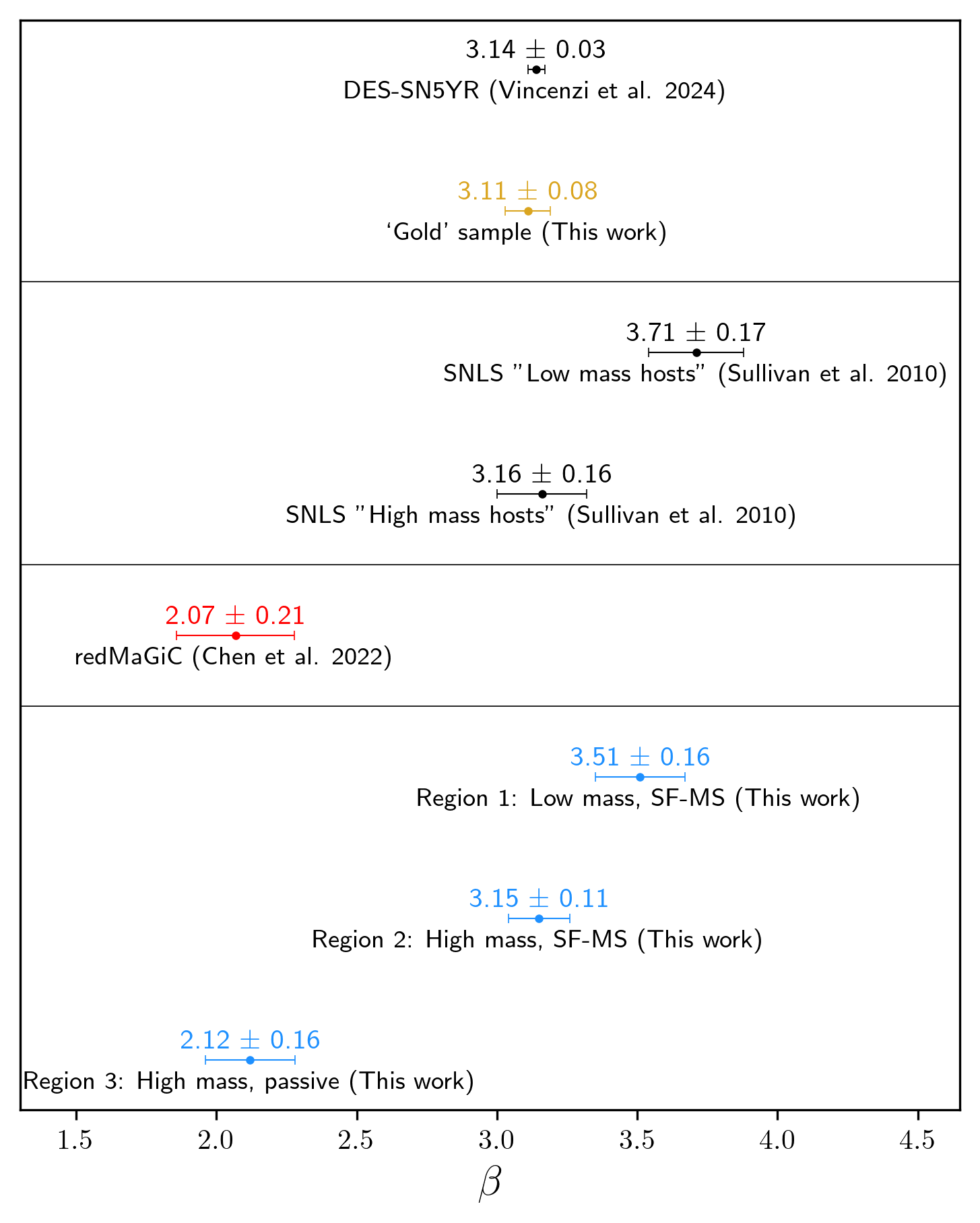}
    \caption{A comparison of values of $\beta$ constrained from select analyses in the literature and this work (at 3$\sigma$ significance).}
    \label{fig:beta}
\end{figure}

We now step through and compare our findings with the relevant literature. Specifically, previous works that have split the SN Ia sample by a selected property of the host galaxy in order to study the impact on the SN colour -- luminosity relation. To aid interpretation, a comparison of $\beta$  from the various analyses listed in the remainder of this subsection are summarised in Fig.~\ref{fig:beta}.

\subsubsection{Splitting by stellar mass}

It has been commonplace in past research (\citealp{2010ApJ...715..743K,2010MNRAS.406..782S}) to use stellar mass as a proxy when studying correlations between SNe Ia and their host environments. The reason being that it is the most accessible host property for which estimates from SED fitting are robust (Fig. \ref{stellarmass}). Using the Supernova Legacy Survey (SNLS) sample, \cite{2010MNRAS.406..782S} explore differences in SN Ia properties when splitting their sample by host mass, testing different thresholds when defining their low- and high-mass galaxy subsamples. When splitting at log($M_\star/M_\odot$) = 10 and determining cosmological nuisance parameters on either side of the split, they find $\beta = 3.71 \pm 0.17$ and $\beta = 3.16 \pm 0.16$ for low- and high-mass galaxies, respectively. Their low-mass value for $\beta$ is consistent with our region 1, which applies the same cut on mass. Regarding their high-mass subsample, this is likely to include both star-forming and passive hosts. Here, their measurement of $\beta$ can be understood in the context of Fig. \ref{beta_vs_sigma}, where $\beta$ for region 2 remains consistent with ${\sim}3.15$ even as the $\sigma$-cut that is used to divide massive hosts is increased. At the 5$\sigma-$cut extreme, it is highly probable that there is contamination from passive hosts in the SF sample. Yet, we do not see a significant reduction in the measured $\beta$, suggesting the star-forming host galaxies have a larger effect on the value of $\beta$. This is likely due to the larger variation in SN colour in star-forming host galaxies, meaning that the more extreme SN colours dictate the slope of the correlation, e.g. Fig.~\ref{betaplot2}. The analysis by \cite{2010MNRAS.406..782S} has since been replicated on much larger and statistically robust datasets, with many consistently recovering the trend of a decreasing $\beta$ with increasing stellar mass (e.g., \citealp{2021MNRAS.508.4656G, 2021MNRAS.501.4861K, 2025A&A...694A...5P, 2025A&A...694A...4G}). 

\subsubsection{Splitting by sSFR}

In order to  more definitively separate galaxies comprised of young and old stellar populations, \cite{2010MNRAS.406..782S} also explore the approach of splitting the sample at different thresholds of specific SFR (sSFR). They observe a significant difference in $\beta$ values between low- and high-sSFR galaxies ($\beta=2.88\pm0.16$ and $3.73\pm0.16$, respectively, when splitting at log(sSFR/yr$^{-1}$) = $- 9.30$). In estimating host galaxy parameters prior to creating these subsamples, \cite{2010MNRAS.406..782S} use optical and NIR photometry, specifically the \textit{ugrizJHK$_\mathrm{s}$} bands on the Canada–France–Hawaii Telescope. However, the absence of mid- and far-infrared data when fitting a model to a galaxy SED can significantly affect SFR (Section \ref{results}); any systematic errors on which naturally propagate to errors in sSFR. When mapping galaxies on the SFR-$M_\star$ plane, this can lead to the region they occupy to vary drastically. To illustrate this, we recreate Fig. \ref{sfms} but in the absence of \textit{Herschel} and \textit{Spitzer} when deriving host galaxy parameter estimates; we present this in Fig. \ref{NHSsfms}. This highlights the misclassification of passive galaxies as actively star-forming hosts (and vice versa) -- a consequence of the age-metallicity-dust degeneracy. Therefore, it is highly probable that the division of galaxies by sSFR in \cite{2010MNRAS.406..782S} is not resulting in the `purest' samples and instead, the low- and high-sSFR groups contain a combination of both passive and star-forming hosts. This has the potential to impact the measurement of $\beta$, as well as other cosmological parameters, potentially increasing contributions to the cosmological systematic error budget. 
We note that the trend we find that is consistent with \cite{2010MNRAS.406..782S} (and other similar works that sub-sample SNe by the star-formation activity of their hosts e.g., \citealp{2010ApJ...722..566L,Chen_2022}) is that SNe in passive galaxies show a preference for lower values of $\beta$ than those in actively SF hosts. As a final point, we mention that \cite{2010MNRAS.406..782S} use a split at a single value of sSFR. We suggest a more robust way to divide galaxies into populations characterised by their star-formation activity is as done in this paper. That is, a division based on where a galaxy lies with respect to the SF-MS (Eq. \ref{whit}).

\subsubsection{Splitting by host galaxy type} \label{redmag_chen}

A distinct alternative to property-based subdivisions (e.g., sSFR, stellar mass) is to construct sub-samples of SNe Ia according to host galaxy type. This approach comes with certain advantages. For example, selecting galaxies of the same mass/age/metallicity irrespective of redshift acts to dilute the effect of any evolutionary trends in host properties across cosmic time. This in turn results in a sample that is less influenced by various empirical correlations between SN light-curve parameters and their galaxy environments.

\cite{Chen_2022} conduct a study of a sample of DES SNe Ia that are located exclusively in Luminous Red Galaxies (LRGs; see \citealp{2001AJ....122.2267E}), which they identify using the the red-sequence Matched-filter Galaxy (redMaGiC; \citealp{Rozo_2016}) algorithm (hereafter referred to as the "redMaGiC" sample). One aspect of their analysis focuses on the SN colour-luminosity relation, for which the redMaGiC sample yields $\beta = 2.068 \pm 0.210$ -- a value that is significantly lower than those reported in the literature for the full SN Ia sample, i.e. without sub-division by galaxy type. The authors propose that this low $\beta$ (or, alternatively, a weaker correlation between SN colour and luminosity) is indicative of a population that is more standardizable and thus, better optimised for use in a cosmological context (particularly when lacking spectroscopic coverage).

\begin{figure*} 
  \begin{subfigure}[b]{0.49\linewidth}
    \centering
    \includegraphics[width=1\linewidth]{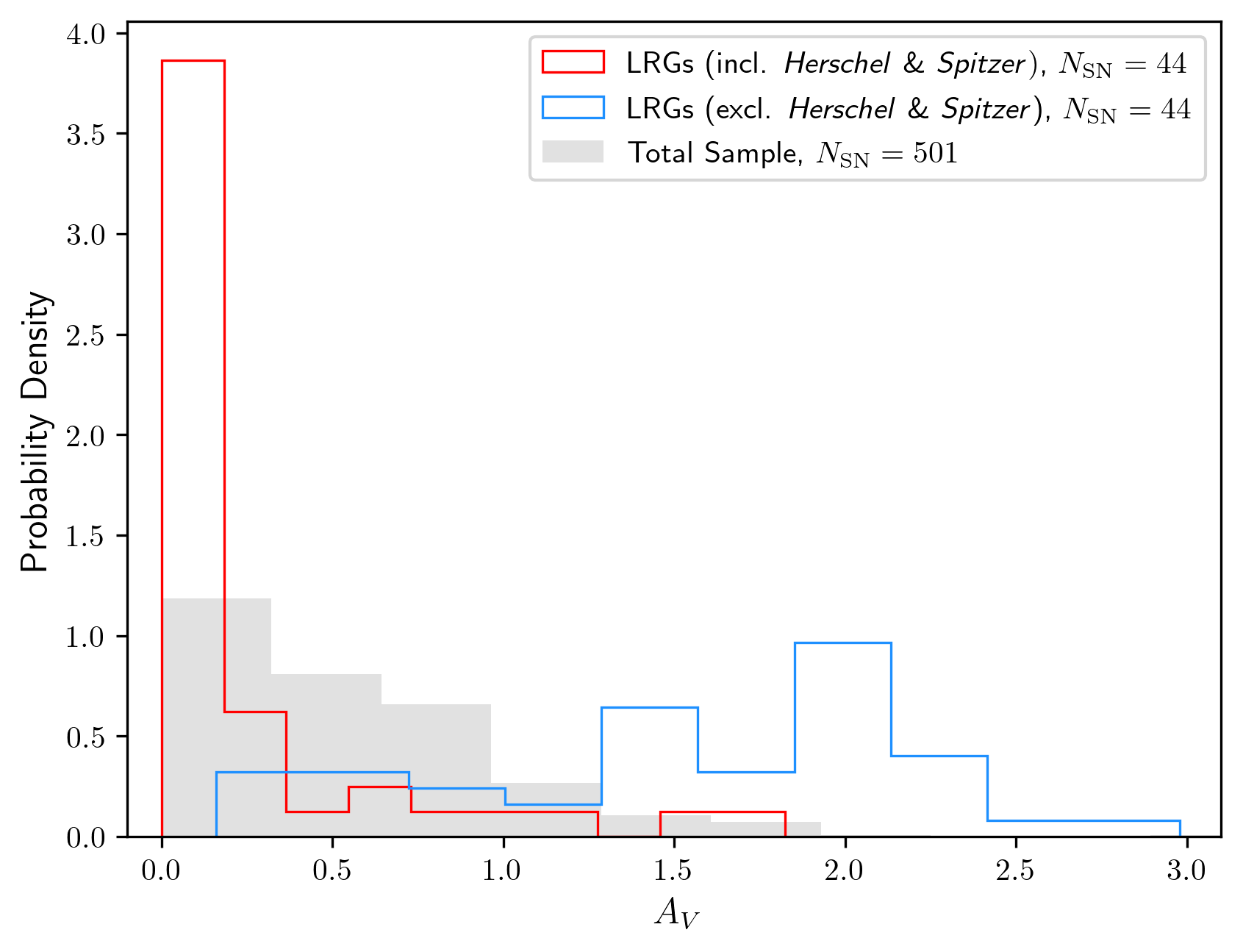} 
  \end{subfigure} \hfill
  \begin{subfigure}[b]{0.49\linewidth}
    \centering
    \includegraphics[width=1\linewidth]{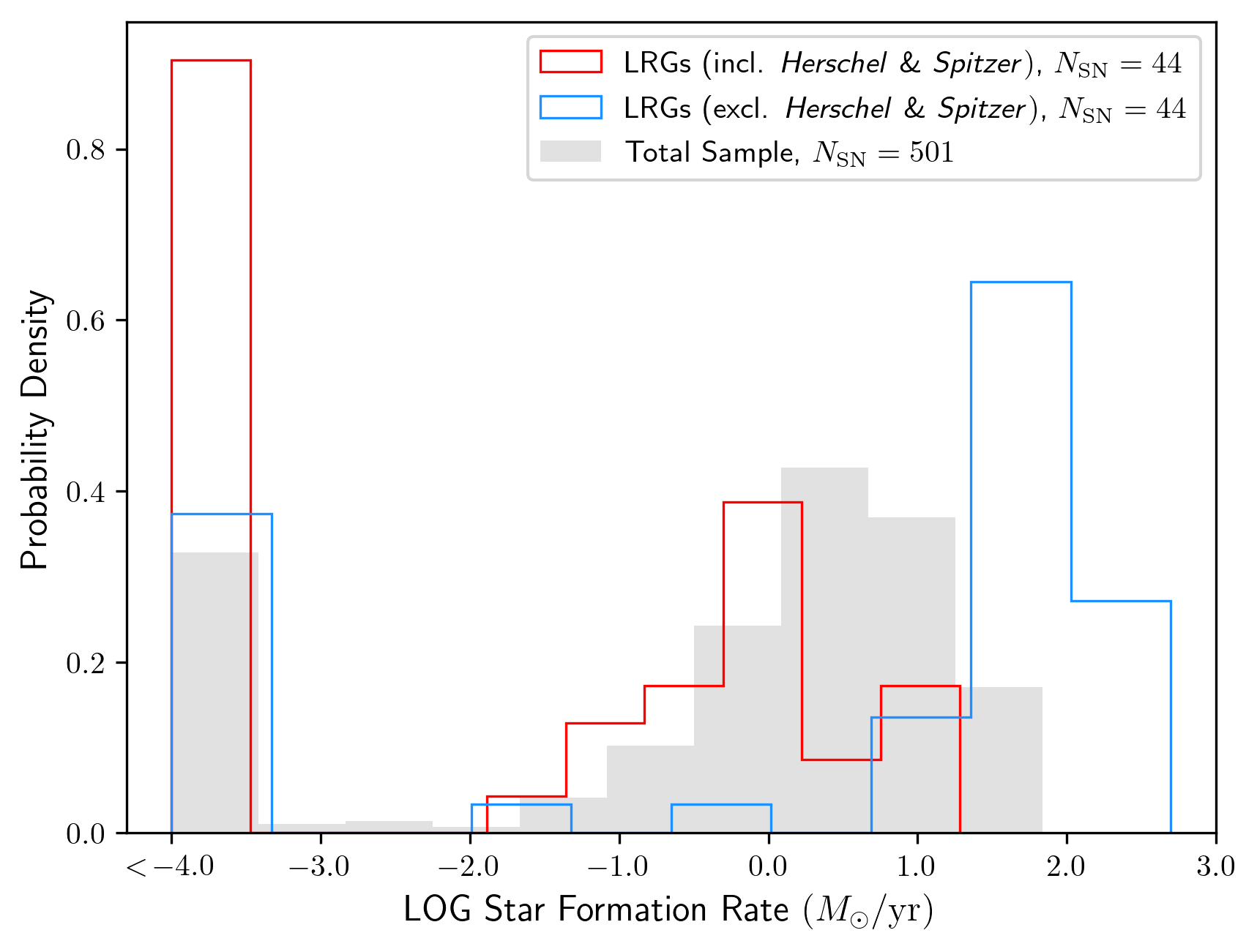} 
  \end{subfigure}
  \caption{Left: The $A_V$ distributions of SN Ia host galaxies for the following cases: (i) redMaGiC LRG sample when using \textit{Herschel} and \textit{Spitzer} data to obtain parameter constraints (red histogram), (ii) redMaGiC LRG sample when excluding \textit{Herschel} and \textit{Spitzer} data in analysis (blue histogram). The sole purpose of including this in the figure is to emphasise one of the key takeaways of this paper, that is, the absence of far-infrared data can result in significantly different host galaxy parameter estimates. (iii) total distribution of entire sample of SN hosts for which we have host galaxy photometry for i.e. our `gold' sample (grey histogram). We note that the axis is truncated solely for visualisation purposes to exclude a single point with $A_V > 3$ mag. Right: As Left but for SFR instead of $A_V$ where every object that has a log SFR < -- 4 $M_\star/\mathrm{yr}$ (extremely passive) is placed into the -- 4 bin.} 
  \label{REDMAGIC} 
\end{figure*}

With our improved host galaxy parameter constraints and an updated catalogue from \cite{Chen_2022} consisting of 212 LRG SN hosts (44 of which we have host galaxy photometry for across CDFS and XMM), we look at what portion of the SFR--$M_\star$ plane is occupied by LRGs. Overplotting the redMaGiC sample hosts on Fig. \ref{sfms}, which we highlight using red contours, we find that LRGs are mainly associated with region 3. For this region, we measure $\beta = 2.12 \pm 0.16$, a value consistent with that obtained by \cite{Chen_2022}. In addition, we present the $A_V$ and SFR distributions of the redMaGiC LRGs in Fig.~\ref{REDMAGIC}, both with and without \textit{Herschel} and \textit{Spitzer} data. The results of interest inform us that LRGs are characterised by little amounts of dust attenuation, with the majority of the sample having $A_V \lesssim 0.3$ mag, and SF, as expected of early-type galaxies that are comprised of old, passive stellar populations.

Finally, we note that the redMaGiC algorithm used to obtain a sample of LRGs in \cite{Chen_2022} only returns a fraction of the objects that populate Fig. \ref{sfms}, region 3. We thereby suggest that, as done here, future analyses use SED fitting of multi-wavelength data to derive a larger and more \textit{complete} sample of hosts, where possible.

\subsection{Mass step}

Traditionally in SN Ia cosmological analyses, the `mass step' is defined as the difference in the average Hubble residuals between low- and high-mass galaxies, when a single set of nuisance parameters (e.g., $\alpha, \beta$ etc.) are fit to the full SN Ia sample. In this analysis, we first split our SN Ia sample into three sub-samples based on the region of the SFR-$M_\star$ plane their host galaxies occupy. We then minimise the residuals and constrain nuisance parameters ($\alpha, \beta, M$) for each sub-sample separately. Therefore, we are unable to investigate the size of the `step' as has been done previously, as each sub-sample has different nuisance parameter constraints. Instead we investigate differences in the SN Ia rest-frame absolute magnitude across each region, which can be thought of as analogous to the traditional `mass step'. Our findings (Table \ref{beta-vals}) are in line with previous works where we find SNe in more massive galaxies are brighter, post-standardization, compared to less massive galaxies. Amongst both our high mass regions (SF-MS vs. passive hosts), we find the brightest SNe are in passive galaxies (region 3). 

As for the origin of the `mass step', the model presented by \cite{2021ApJ...909...26B} implies that it is an artefact of different $\beta$ in hosts of different mass and that by accounting for a different slope of the dust extinction law $R_V$, and thus effectively correcting SNe by different $\beta$, the step can be removed. However, other works have found a residual mass step even after accounting for independent fitting of $\beta$ and $M$ in different hosts \citep{2010MNRAS.406..782S, 2011ApJ...737..102S}, as well as specifically modelling $R_V$-like effects \citep{Wiseman2022, 2023MNRAS.519.3046K}. The DES-SN5YR analysis was performed allowing for maximal $R_V$ variation between low- and high-mass galaxies, following \cite{2021ApJ...913...49P}, but found a 0.04 mag residual mass step \citep{2024ApJ...975...86V}. Our results are similar: we find that even when correcting SNe in each sub-sample for their `best' constrained nuisance parameters, the differences in their luminosities persist. 

We highlight that $\beta$ appears to decrease when moving from young, actively star-forming environments, along the main-sequence and then eventually dropping off it, tracking an evolution in the average progenitor age. The `mass step' (or, perhaps rather, `region step') that remains may thus simply be reflective of intrinsic differences between SN Ia progenitors as a function of age. This is endorsed by a growing body of literature that explores the effects of galaxy/progenitor age on SN Ia standardisation (e.g., \citealp{2013A&A...560A..66R,2018ApJ...854...24K,2018A&A...615A..68R,2019ApJ...874...32R, 2020A&A...644A.176R, 2021MNRAS.501.4861K,2022A&A...657A..22B, Wiseman2022, 2023MNRAS.520.6214W, popovic_modelling_2024}). Using simulations, \citet{Wiseman2022} find that a combination of galaxy-age-driven dust and an intrinsic SN progenitor age luminosity step best reproduce the observed Hubble-residual -- SN colour trends. \cite{2020A&A...644A.176R} split their SN Ia sample by the local sSFR (lsSFR), which they derive from H$\alpha$ flux within a projected 1 kpc radius around each SN location. They find the significance of the step is greater for lsSFR, which is a better proxy for progenitor age than stellar mass, with an amplitude of $0.163 \pm 0.029$ mag $(5.7\sigma)$. \cite{2022A&A...657A..22B} also show that the step size directly correlates with the ability of a host galaxy parameter to trace the stellar population age local to the SN. In light of this evidence, we recommend that understanding the influence of progenitor age on SN Ia luminosities remain central to future investigations.

\section{Conclusions} \label{conclusion}

The principle findings from our investigation are as follows, 

\begin{enumerate}[label=(\roman*), leftmargin=*,align=left,labelindent=\parindent, labelsep=0.5em, itemindent=!]
    \item The inclusion of mid- and far-infrared data from the \textit{Herschel} and \textit{Spitzer} space telescopes in SED fitting can result in host galaxy parameter estimates of dust attenuation and SFR that differ significantly when longer wavelength data is excluded. Mass estimates using only optical/NIR data are generally robust. 
    \item Host galaxy $R_V$ is largely unconstrained even by our extensive multi-wavelength dataset.
    \item The slope of the SN colour-luminosity relation ($\beta$) is dependent upon the region of the SFR-$M_\star$ occupied by their host galaxies. SNe in high-mass, passive galaxies yield the lowest value of $\beta = 2.12 \pm 0.16$ ($>6\sigma$ difference compared to region 1), and show a smaller r.m.s scatter in the Hubble residuals.
    \item SNe Ia in high-mass, passive galaxies are $0.07 - 0.12$ mag ($>3\sigma$) brighter post-standardisation than their low-mass and high-mass, star-forming counterparts. This finding comes after applying separate corrections to SN stretch and colour based on where their host galaxies lie on the SFR-$M_\star$ plane (as opposed to a global correction, as is usually done in cosmological analyses).
    \item We recover previous trends that high-mass galaxies host SNe Ia with narrower light-curve widths (`stretch'), with passive hosts driving stretch distributions to lower values of $x_1$. No significant differences are observed in SN colours based on their host galaxy environments. The mean of the colour distribution displays a slight shift toward redder values for SNe in high-mass, star-forming hosts.
    \item Going forward, cosmological analyses using SNe Ia should identify where on the SFR-$M_\star$ plane the host galaxies of all SNe lie. This will enable sub-samples of SNe Ia to be isolated for use in a cosmological context, where the relevant corrections to SN colour and stretch are applied, depending on their host galaxy environments.
    \item Alternatively, in the interest of higher number statistics, cosmological analyses may wish to use the full SN Ia sample. In this case, the intuitive approach would be to correct SNe in each of the three regions by their respective nuisance parameters, before combining them onto a single Hubble diagram to allow for cosmological parameter constraints. This approach, however, introduces additional nuisance parameters in the cosmological analysis framework (e.g., three separate $\beta$ corrections). Therefore, prior to making any conclusive remarks, a thorough model comparison analysis would first have to be done -- using metrics such as the Akaike/Bayesian information criterion, for example -- to assess whether the increased degrees of freedom significantly improve constraints on cosmological parameters e.g., $w, \Omega_M$, to justify their use. This should form the basis of future work.
\end{enumerate}

This work highlights the importance of including mid-- to far--infrared data, specifically from the \textit{Herschel} and \textit{Spitzer} space telescopes, in breaking degeneracies when deriving select galaxy parameters. With future time domain surveys, such as the Rubin Observatory Legacy Survey of Space and Time (LSST; \citealp{2019ApJ...873..111I}), set to bring in observations of SNe Ia on orders of magnitudes larger, we must now consider the feasibility of performing such an analysis on much more extensive host galaxy datasets. LSST, in particular, will map $\sim$18,000 deg$^2$ of the southern hemisphere in six broadband optical filters (\textit{ugrizy}). Therefore, for studies of (but not limited to) SN hosts, it will be necessary to supplement the optical data with longer wavelength ancillary datasets from other sky surveys. Of the total LSST observing footprint, deep \textit{Herschel} and \textit{Spitzer} data is only available over the Deep Drilling Fields (DDFs) -- smaller regions ($\sim$ 100 deg$^2$ total) of sky that are set to be sampled to increased cadence and depth, although data from the Wide-field Infrared Explorer \citep[WISE; ][]{WISE} may allow some work to be done to mid-infrared wavelengths at lower redshift over larger areas. This naturally raises the question, for regions of the sky covered by the LSST (or any future time domain survey) observing footprint that do not have complementary mid- and far-infrared data (not necessarily from \textit{Herschel} and \textit{Spitzer}), will we be able to obtain robust constraints and break degeneracies between various host galaxy parameters? This will be of vital importance going forward as the ability to standardize SNe Ia for precision cosmology rests on knowing what region of the SFR-$M_\star$ plane their host galaxies occupy, so that the appropriate corrections can be applied to their colour and stretch.

\section*{Acknowledgements}

The authors thank Lisa Kelsey, Brodie Popovic and Dan Scolnic for useful discussions and providing extensive feedback. We also thank the anonymous referee for their careful reading of the manuscript and providing detailed comments. SR, MV and MJJ acknowledge support from the Oxford Hintze Centre for Astrophysical Surveys which is funded through generous support from the Hintze Family Charitable Foundation. MJJ also acknowledges the support of a UKRI Frontiers Research Grant [EP/X026639/1], which was selected by the European Research Council. PW acknowledges support from the Science and Technology Facilities Council (STFC) grant ST/Z510269/1.

\section*{Data Availability}

The optical and infrared data used to measure the host galaxy properties are all in the public domain. The DES-SN5YR data are all publicly available \citep{2024ApJ...975....5S}\footref{desdr}.



\bibliographystyle{mnras}
\bibliography{example} 




\appendix

\section{SExtractor parameters used for Spitzer IRAC catalogue creation} \label{A}

\begin{table}
	\centering  
	\caption{SExtractor input parameters for \textit{Spitzer} IRAC channels 1 and 2. \textsuperscript{\textdagger}Coverage maps are used as inverse variance weight maps.}  
	\label{sex}  
	\begin{tabular}{lc} 
		\hline  
		\hline  
		 Parameter & Value\\  
		\hline  
		  DETECT$\_$MINAREA & 3 \\  
		  DETECT$\_$THRESH & 1.4 \\  
		  ANALYSIS$\_$THRESH & 1.4 \\  
		  FILTER & N \\  
          DEBLEND$\_$NTHRESH & 64 \\
          DEBLEND$\_$MINCONT & 0.001 \\
          SEEING$\_$FWHM & 1.8 \\ BACK$\_$SIZE & 32 \\ BACK$\_$FILTERSIZE & 5 \\ BACKPHOTO$\_$TYPE & LOCAL \\ WEIGHT$\_$TYPE & MAP$\_$WEIGHT\textsuperscript{\textdagger} \\
		\hline  
	\end{tabular}  
\end{table}

\section{\textsc{Bagpipes} SED fits and posterior distributions} \label{appendix:B}

    
    
    

\begin{figure*}
    \vspace*{\fill}
    \centering
    \begin{subfigure}[b]{1\linewidth}
        \centering
        \includegraphics[width=\linewidth]{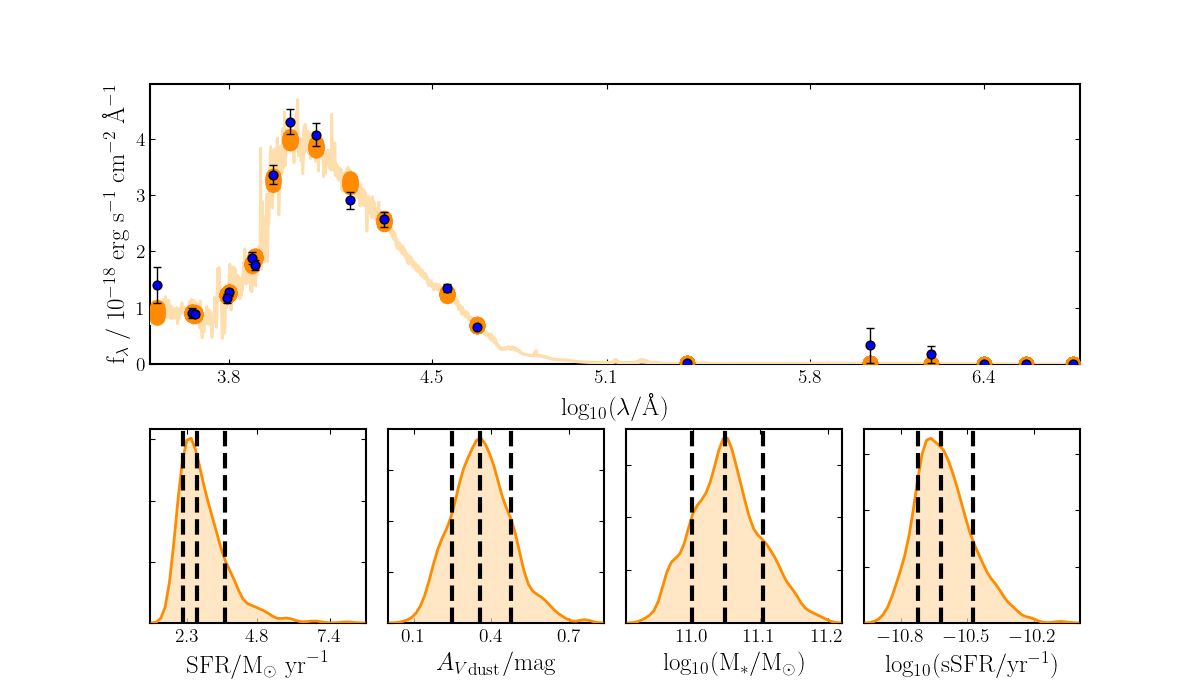}
    \end{subfigure}
    \label{fig:bagcorner1}
    \caption{\textsc{Bagpipes} SED model fits to DES SN host galaxy (CID 1313284) photometry and galaxy parameter posterior distributions.}
\end{figure*}

\begin{figure*}
    \vspace*{\fill}
    \ContinuedFloat
    \centering
    \begin{subfigure}[b]{1\linewidth}
        \centering
        \includegraphics[width=\linewidth]{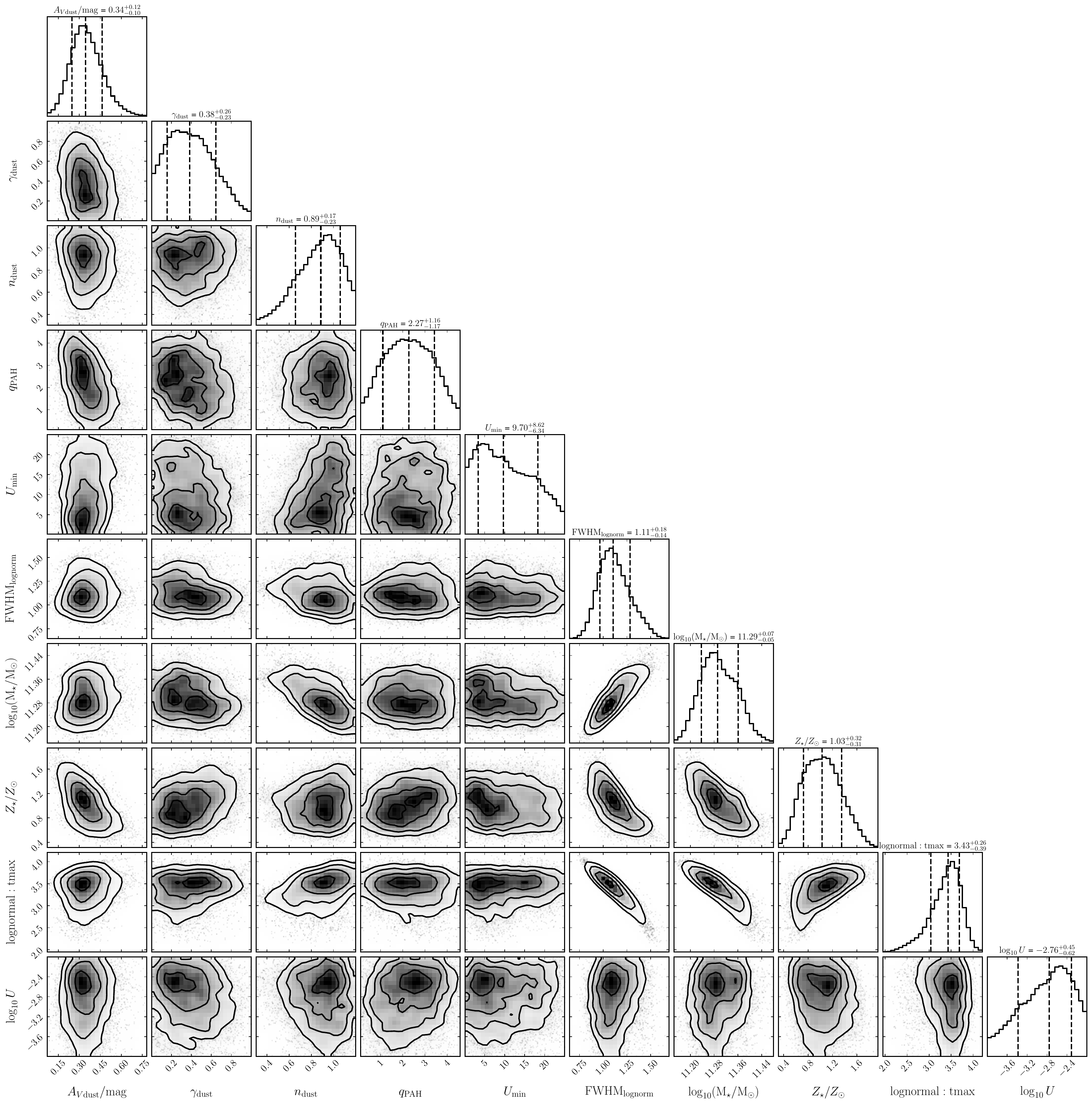}
    \end{subfigure}
    \caption*{(Continued from previous page)}
\end{figure*}



\section{Alternative representations of the SFR-$M_\star$ plane}

\begin{figure*}
    \centering
    \includegraphics[width = 178mm]{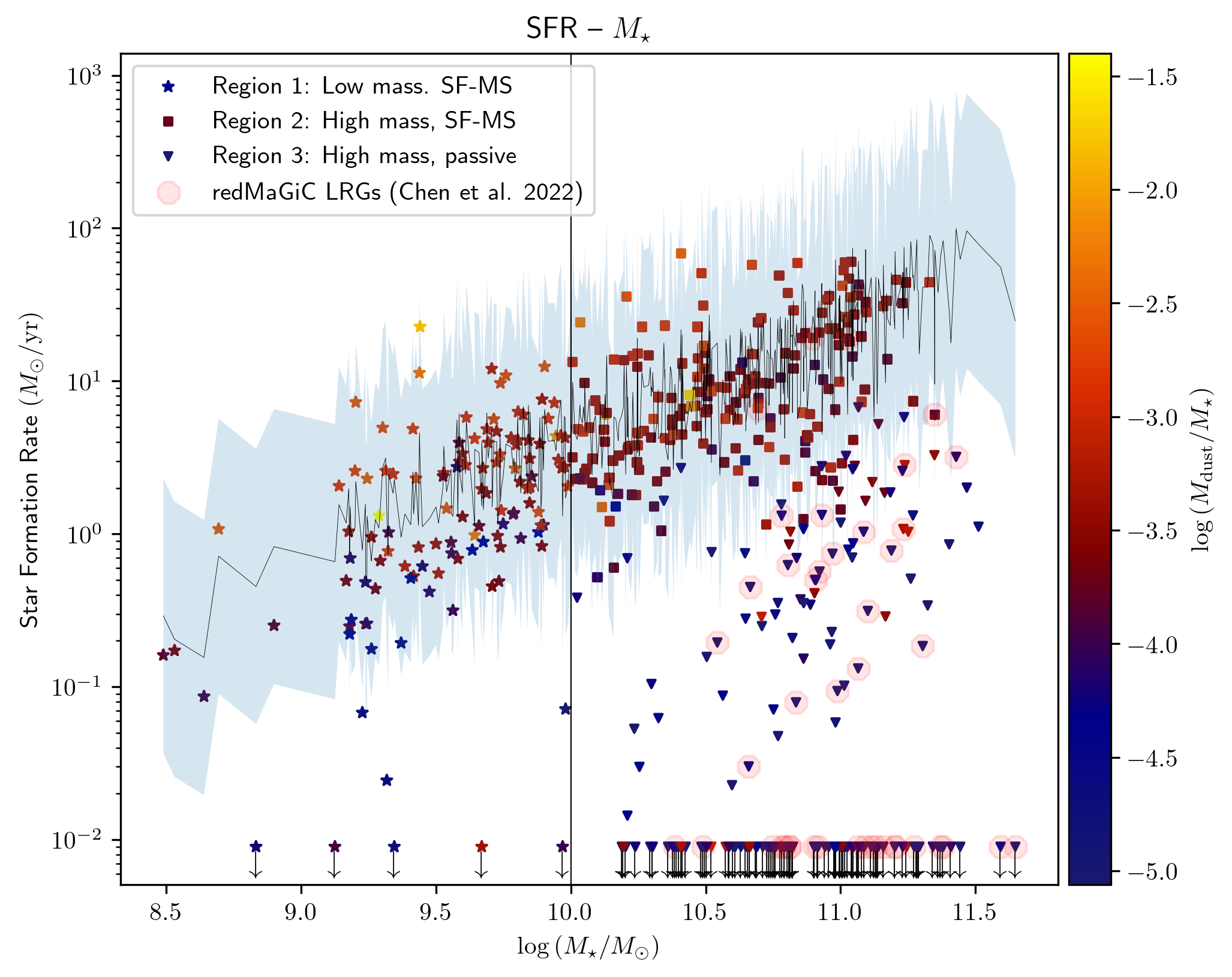}
    \caption{As Fig. \ref{sfms} (i.e., host galaxy parameter constraints are derived using \textit{Herschel} and \textit{Spitzer}) but colour-mapped by log specific dust mass. The blue shaded region represents a 3$\sigma$ interval with respect to the SF-MS. The dynamic range on the colour bar is altered to limit the lower bound on $\mathrm{log}(M_{\mathrm{dust}}/M_\star)$
    to -5, to emphasise differences between specific dust mass across
    the SFR – $M_\star$ plane.}
    \label{Mdsfms}
\end{figure*}

\begin{figure*}
    \centering
    \includegraphics[width = 178mm]{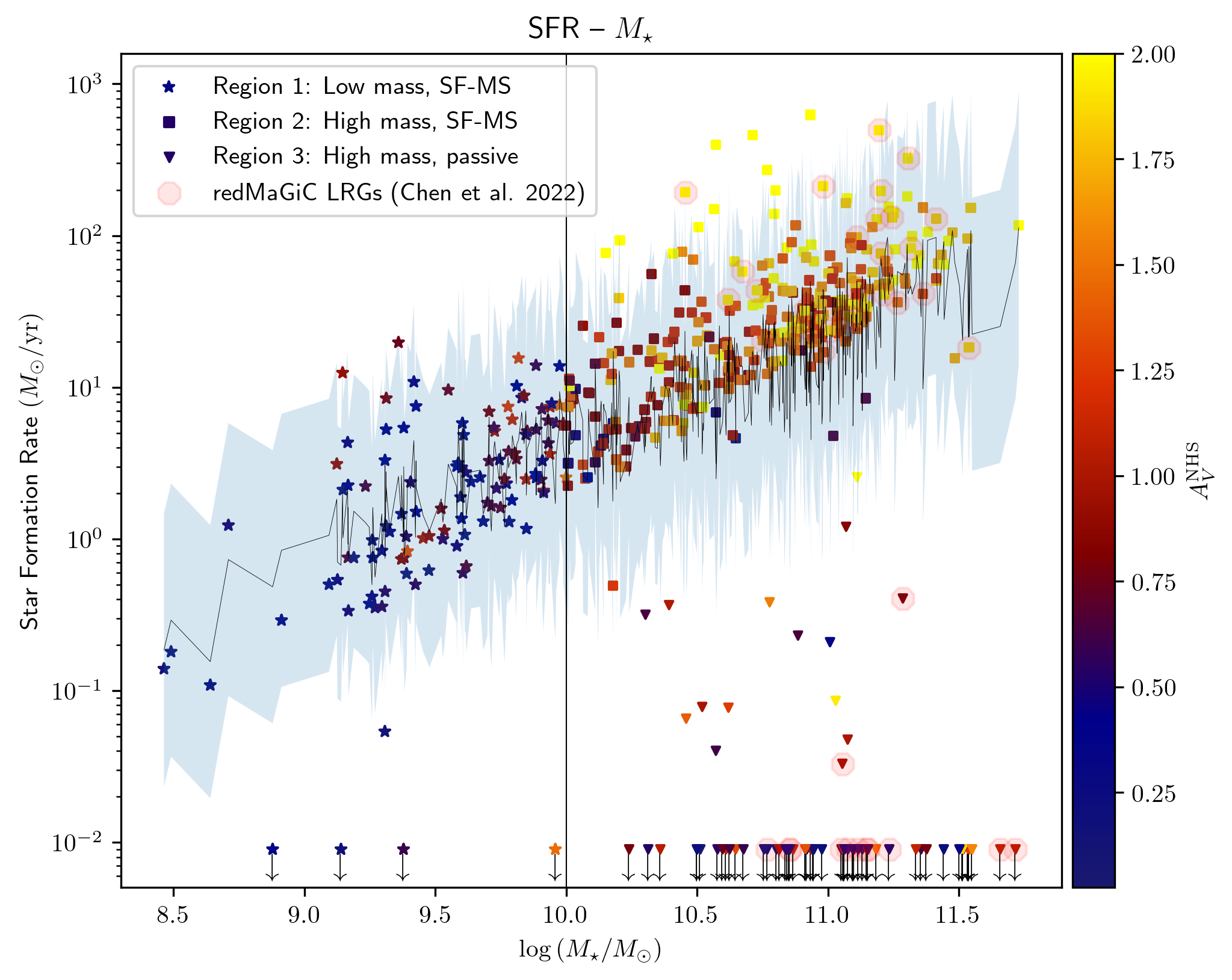}
    \caption{As Fig. \ref{sfms} but for host galaxy parameters derived when \underline{\textbf{excluding}} \textit{Herschel} and \textit{Spitzer}. The blue shaded region represents a 3$\sigma$ interval with respect to the SF-MS. The galaxies are mapped as a function of dust attenuation estimates obtained in the absence of \textit{Herschel} and \textit{Spitzer} data, $A_V^{\mathrm{NHS}}$. Colour-mapping by $A_V^{\mathrm{NHS}}$ illustrates the findings in Fig. \ref{res_dust}, with the dust attenuation incorrectly estimated for a large fraction of the population.}
    \label{NHSsfms}
\end{figure*}


\bsp	
\label{lastpage}
\end{document}